\documentclass[journal]{IEEEtran}
\usepackage{graphicx}
\usepackage{caption}
\usepackage{subcaption}
\usepackage{url}

\usepackage{booktabs}
\usepackage{amsmath}
\usepackage{xcolor}

\newcommand\dblquote[1]{\textquotedblleft #1\textquotedblright}

\ifCLASSINFOpdf
\else
\fi
\begin{document}
	\title{NEMA: Automatic Integration of Large Network Management Databases}
	
	\author{Fubao Wu,  %~\IEEEmembership{Member,~IEEE,}
		Han Hee Song, %~\IEEEmembership{Fellow,~OSA,}
		Jiangtao Yin,
		Lixin Gao,
		Mario Baldi,
		Narendra Anand %~\IEEEmembership{Life~Fellow,~IEEE}% <-this % stops a space
		\thanks{Fubao Wu and Lixin Gao are with the Department
			of Electrical and Computer Engineering, University of Massachusetts, Amherst. E-mail: fubaowu@umass.edu}% <-this % stops a space
		\thanks{Han Hee Song, Jiangtao Yin, Mario Baldi and Narendra Anand are with Cisco.}% <-this % stops a space
	}

	\maketitle
	
	\begin{abstract}
		Network management, whether for malfunction analysis, failure prediction, performance monitoring and improvement, generally involves large amounts of data from different sources. To effectively integrate and manage these sources, automatically finding semantic matches among their schemas or ontologies is crucial. Existing approaches on database matching mainly fall into two categories. One focuses on the schema-level matching based on schema properties such as field names, data types, constraints and schema structures. Network management databases contain massive tables (e.g., network products, incidents, security alert and logs) from different departments and groups with nonuniform field names and schema characteristics. It is not reliable to match them by those schema properties. The other category is based on the instance-level matching using general string similarity techniques, which are not applicable for the matching of large network management databases. In this paper, we develop a matching technique for large NEtwork MAnagement databases (NEMA) deploying instance-level matching for effective data integration and connection. We design matching metrics and scores for both numerical and non-numerical fields and propose algorithms for matching these fields. The effectiveness and efficiency of NEMA are evaluated by conducting experiments based on ground truth field pairs in large network management databases. Our measurement on large databases with 1,458 fields, each of which contains over 10 million records, reveals that the accuracies of NEMA are up to $95\%$. It achieves $2\%$--$10\%$ higher accuracy and 5x--14x speedup over baseline methods.
		
	\end{abstract}
	

	% \fi
	\IEEEpeerreviewmaketitle
	
\section{Introduction}
\label{Introduction}
	\IEEEPARstart{W}{ith} the development of big data analytics and data mining techniques,  scalable network measurement and analysis techniques have been used in finding hidden information or patterns to help with network management, network monitoring and network security \cite{wang2016big,dastbaz2015green, himmelstein2015heterogeneous, sahu2017ubiquity}. In one scenario of important network management, big network vendors serving various customers own a multitude of databases on network product information, network configurations, geographical information on deployment sites, incident logs, etc. Since the databases serve different departments, they are usually separated from each other and independently managed by different departments and groups. However, network data are inherently designed to host ``connections'' among different devices, groups that the devices are in, and functionality the devices serve together; they are often required to be shared and put together to be used in many important tasks such as network prediction, semantic query, network analysis and fault detection. For example, with the data connections, one can easily model a graph database to query related product configurations and performance issues around faulty devices or service incidents, which are matched with another product-related table \cite{pantuza2014network, aggarwal2010graph}. As many studies \cite{daraio2016advantages, dong2015big,  chen2018biggorilla, stonebraker2018data} point out, the discovery of matching fields is the most crucial and foremost step for the integration of databases. For this reason, in this paper we aim to automatically construct such matchings that lead to efficient network management and analysis.  
	
	There is an abundance of research on field matching and integration approaches for different data formats in different contexts such as relational databases, XML and object-oriented data formats \cite{gomez2014data,raut2014survey, sunderhauf2017meaningful, castano2018matching}. Existing database matching approaches include two main categories of techniques. One is based on schema-level matching, which exploits metadata using schema characteristics such as field names, data types, structural properties and other schema information \cite{bhattacharjee2009ontomatch, gu2016interaction}. However, network management databases from different sources have different design standards and naming conventions \cite{jang2017history, coronel2016database}. Even similar fields can have different names (e.g. ``product\_family'' can also be named as ``product series''). The other category is instance-level matching, which uses the record values of two fields to calculate the similarity and to determine matched fields \cite{chen2012mining, qian2012sample, nottelmann2007information, mehdi2012instance}. Most of the previously proposed schemes rely on syntactic similarities, sampling or machine learning techniques that are meant to extract common patterns from the matching data corpus. However, it is difficult to directly apply these techniques to network databases or challenging to reliably construct dictionaries, corpus, and training labels with large datasets when the naming convention is not consistent and diverse.
	
	The challenges for matching these network databases are:  (1) The database design is not ideally uniform. The data tables are created in different groups and departments by different people. Therefore, it is not reliable to use schema information to match data directly and easily. (2) The data is noisy and irregular. Some table fields contain unexpected records such as null, invalid values and typos. Some table records are either partly missing, incomplete or incorrect. Some fields have a large amount of records, while some have very few records. (3) The table contents are complicated and heterogeneous with numerical and non-numerical data formats. (4) No thesauri or auxiliary information exist that we can rely on for matching. The only set of observation available is the database itself.     
	
	To solve these challenges, we propose an automatic matching technique for large NEtwork MAnagement databases (NEMA) to construct a graph database for network management and analysis effectively and efficiently. We propose several algorithms for numerical matching and non-numerical respectively based on instance matching. 
	Our main contributions are as follows:
    \begin{enumerate} 
    \item We propose effective range difference similarity and bucket dot product similarity metric to match numerical fields, and top priority metric to match non-numerical fields.
    \item To make the algorithm more scalable for large network management databases, we utilize min-hash-locality sensitive hashing algorithm for faster processing with a little scarification of accuracy.
    \item We further propose to use the proposed similarity metric scores as features for classification to improve the reliability of our matching technique.
    \item We experimentally demonstrate the effectiveness and efficiency of our matching algorithms in the real Cisco network management databases for constructing a graph network database.
	\end{enumerate}

	The rest of this paper is organized as follows. We define the problem in Section \ref{ProblemDescription}. Section \ref{MatchingAlgorithm} describes NEMA matching algorithms in detail including both numerical matching and non-numerical matching. Experimental evaluation is presented in Section \ref{ExperimentalEvaluation} and related work is shown in Section \ref{RelatedWork}. We conclude our paper in Section \ref{Conclusion}.

	\section{Problem Description}
	\label{ProblemDescription}
	Given structured network management databases, our goal is to create a graph database of network management by finding the most accurate matched field pairs among different tables in these databases. The matching of two fields is determined by the matching score measured by their record pair similarities. We utilize the matched results to construct a graph database for semantic query, network analysis, network prediction, etc. \cite{miller2013graph, lissandrini2018beyond, lose2019combat}. 
	
	To illustrate our problem and algorithms clearly, we use three sample tables below as a toy example throughout the rest of this paper. In Table \ref{tbl-tp}, PRODUCT Table ($T_P$) contains 2 fields $product\_id$ (primary key), and $family$ with 7 records respectively. In Table \ref{tbl-ti},  INCIDENT Table ($T_I$) contains 2 fields $incident\_key$ (primary key), and $prod\_key$ with 7 records respectively. In Table \ref{tbl-to},  ORDER Table (
	$T_O$) contains 3 fields $order\_key$ (primary key), $incident\_id$, and $product\_name$ with 7 records respectively. The problem is to find whether these 7 fields match among the Table \ref{tbl-tp}, \ref{tbl-ti} and \ref{tbl-to} by evaluating the matching of their records, then to construct a graph database for efficient network analysis and management.
	
	\begin{table}[!htb]
		%\caption{Global caption}
		\begin{minipage}{.5\linewidth}
			\captionsetup{skip=4pt}
			\caption{PRODUCT ($T_P$)} \label{tbl-tp}
			\begin{tabular}{ll}
				\hline
				{\bf \textit{product\_id}}  & \textit{family}  \\ \hline
				107                   & AIR series  \\
				108                   & con series \\
				109                   & con series \\
				150                   & 47-7000   \\
				151                   & cisco0500  \\
				152                   & 80-7066C  \\
				153                   & con5100 \\ \hline
			\end{tabular}
		\end{minipage}%
		\begin{minipage}{.5\linewidth}
			\captionsetup{ skip=4pt}
			\caption{INCIDENT ($T_I$)} \label{tbl-ti}
			\begin{tabular}{ll}
				\hline
				{\bf \textit{incident\_key}} & \textit{prod\_key}   \\ \hline
				201          & 107    \\
				202          & 107    \\
				203          & 108    \\
				204          & 109    \\
				207          & 150    \\
				208          & 151    \\
				209          & 152    \\ \hline
			\end{tabular}
		\end{minipage} 
		\vspace*{-5pt}
	\end{table}
	
	\begin{table}[!htb]
		\captionsetup{skip=4pt}
		\caption{ORDER ($T_O$)} \label{tbl-to}{%
			\begin{tabular}{lll}
				\hline
				{\bf \textit{order\_key}} & \textit{incident\_id} & \textit{product\_name}   \\ \hline
				301     & 201     & AIR1212AC        \\
				302     & 201     & AIR1002      \\
				303     & 203     & con5122  \\
				304     & 204     & mem-4700m-64d=  \\
				305     & 207     & 47-7066C   \\
				306     & 208     & cisco0510    \\
				307    & 208     & cs6012    \\ \hline
			\end{tabular}}	
			%\vspace*{-13.5pt}
		\end{table}
		
		%	Some important concepts and our problem are defined here. \\
		
		{\bf Record Matching:} Given two instances e1 and e2, a record matching function is defined as a 4-tuple: $\left \langle  e_1, e_2, v, r\right \rangle$ where e1 and e2 are two field records; $v$ is a similarity score (typically in the [0, 1]) between e1 and e2; $r$ is a relation (e.g., equivalence, part-of, overlapping, etc.) between e1 and e2. The matching function $\left \langle  e_1, e_2, v, r\right \rangle$ asserts that the relation $r$ holds between the record $e1$ and $e2$ with score $v$. For numerical matching, only if two records are equal, they are considered matched. For example, records \{107, 108, 109, 150, 151, 152 \} in $product\_id$ field in Table \ref{tbl-tp} are matched as the equal records in $prod\_key$ field in Table \ref{tbl-ti}, respectively. For a non-numerical pair, however, if the similarity score of two records are higher than a threshold based on a similarity metric, they are considered matched. Here we consider the part-of relation in the network management databases for meaningful relations including subgroup versus group, product versus product family, subseries versus series, and so on. For example, the record \dblquote{con5122} in $product\_name$ field and the record \dblquote{con5100} in $family$ field can have high similarity score with part-of relation. A record pair which is matched is called a matched record pair, and it is called a non-matched record pair if the pair is not matched.
		
		{\bf Field Matching:} A field here means a database field indicating the names of a column and the single piece of data stored. Given two fields $f_1$, $f_2$ and a threshold $T$, we define $sim(f_1,f_2)$ as the matching/similarity score (e.g. Jaccard similarity) between two fields $f_1$ and $f_2$.  If $sim(f_1,f_2)$ value is above $T$, we call $(f_1,f_2)$ a matched field pair, otherwise it is called a non-matched field pair. In the toy example, $sim(product\_id, prod\_key)$ has a high matching score with Jaccard similarity, so $(product\_id, prod\_key)$ can be correlated and matched. Moreover, the field pair $(family, product\_name)$ can also be matched in terms of many matched record pairs such as some pairs $(AIR series, AIR1002)$, $(47-7000, 47-7066C)$ and $(con5100, con5122)$, and so on.
		
\begin{figure}[!th]
			\captionsetup{}
			\centering
			\includegraphics[width=3.1in,height=2.1in]{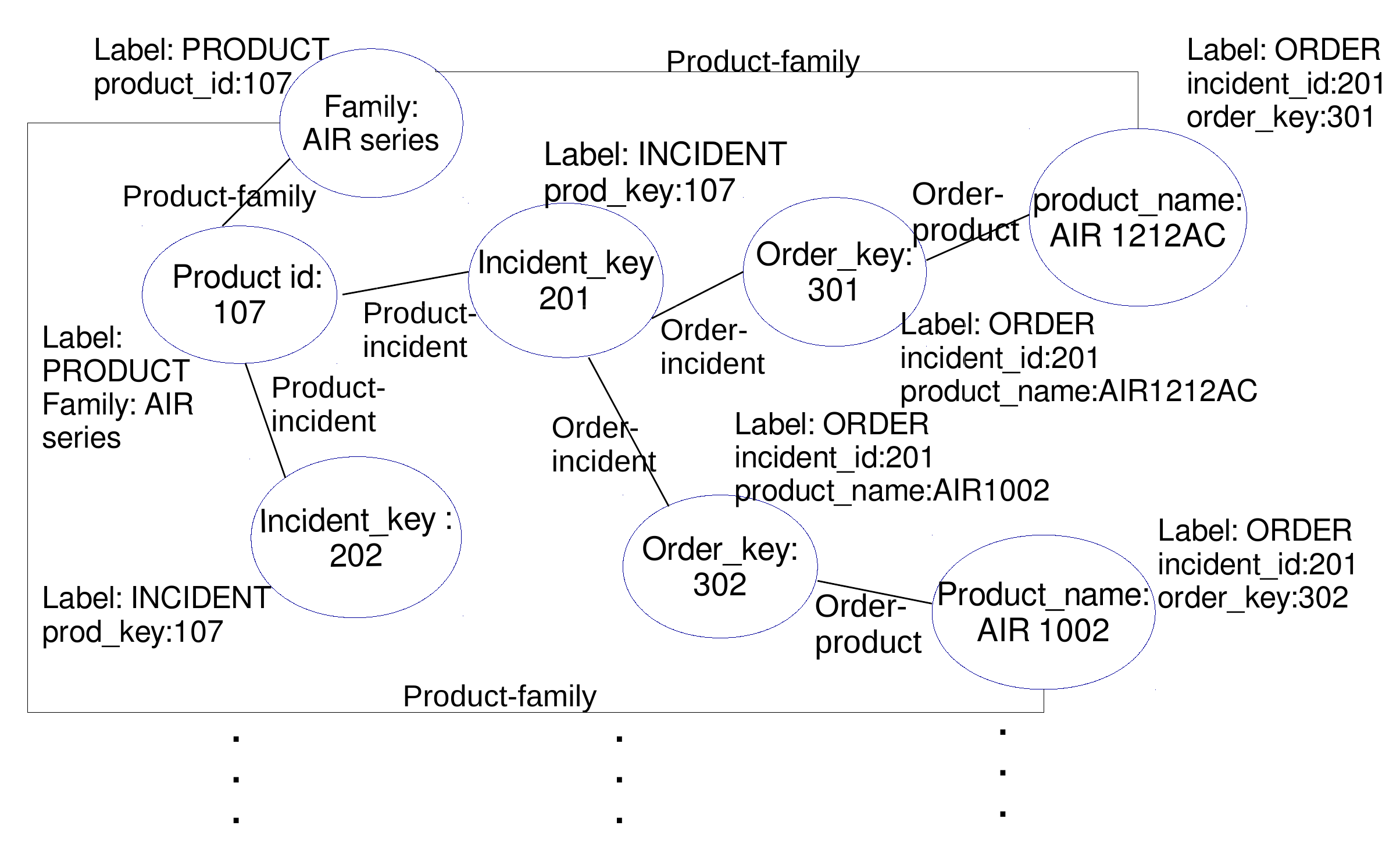}
			\caption{Graph database created from Table  \ref{tbl-tp}, \ref{tbl-ti} and \ref{tbl-to}}
		    	\label{fig:graphDataModelExampleCropped}
		\end{figure}

		{\bf Graph Database:} One effective way to utilize matched results is to construct a graph database for semantic query, network analysis, network prediction, etc. \cite{miller2013graph, lissandrini2018beyond, lose2019combat}, which is also our goal. We define a graph database as a labeled, attributed and undirected graph $G= (V, E, L_v, L_e)$ where $V$ is the node set containing all the records appearing in the fields which match, $E$ is the edge set between node pairs for node set $V$. $L_v$ is a set of label information of node set $V$, which are the index attributes for the columns of the table. $L_e$ is a set of label information of edge sets $E$, which are the relations of two records from two tables' column attributes. Specifically, when we construct a graph database from matching of relational databases, a node $v$ consists of a field and a record value in a row; the label $L_v$ of $v$ comprises of the other field information in the same row as node $v$; an edge $e$ is the matching between two records; $l_e$ indicates the field information when two records of the fields matches.  Figure \ref{fig:graphDataModelExampleCropped} shows an example of a constructed graph database from parts of matched results in Tables \ref{tbl-tp}, \ref{tbl-ti} and \ref{tbl-to}. For example, a specific node $v$ \dblquote{product\_id: 107} in the Table \ref{tbl-tp}  has a node label $l_v$ \dblquote{PRODUCT} and \dblquote{Family:AIR series}. Node $v$ matches with \dblquote{prod\_key: 107} in Table \ref{tbl-ti}, so it is connected to the node \dblquote{Incident\_key:202} with a Product-incident relation, to the node \dblquote{Incident\_key: 201} with a Product-incident relation, and to the node \dblquote{Family:AIR series} with a Product-family relation.
		
		% /home/fubao/workDir/Research/Cisco_WISH/figures_writing_paper.odg
		\vspace{-0.2cm}

		\section{Matching Algorithm}
		\label{MatchingAlgorithm}
		To find whether two fields match, one simple way is to use field name matching. If the name of two fields are the same or similar, they are matched. However, this is not reliable for many database sources because they are noisy and irregular. Moreover, the network databases comprise of numerical and non-numerical fields, and they have different attributes and matching requirements. For numerical matching, we consider equivalence relation between record pairs. For non-numerical matching, however, we do not directly consider the equivalence relation as the matching standard. non-numerical record values are possible to be semantically correlated with different names. For example, the non-numerical fields $family$ and $product\_name$ in the example Table \ref{tbl-tp} and \ref{tbl-to} have very few common characters on their names, but they are semantically correlated that $product\_name$ has a part-of relation with $family$. Moreover, in terms of  field records, records \dblquote{cisco0510} and \dblquote{cisco0500} in these two fields can be considered belonging to the same family and being matched with a high similarity. However, the record pair \dblquote{47-7066C} and \dblquote{80-7066C} are considered to be non-matched with different families, even though the two strings have many common characters. Details will be discussed in \ref{algNonNumericalMatching}). Hence, we make use of the record matching to decide whether two fields match to improve matching accuracy and satisfy semantic matching. Overall, we match numerical and non-numerical fields separately and design different matching algorithms for each of them.
		
		\subsection{System Overview}
\begin{figure}[!th]
			\centering
			\includegraphics[width=3.2in,height=1.2in]{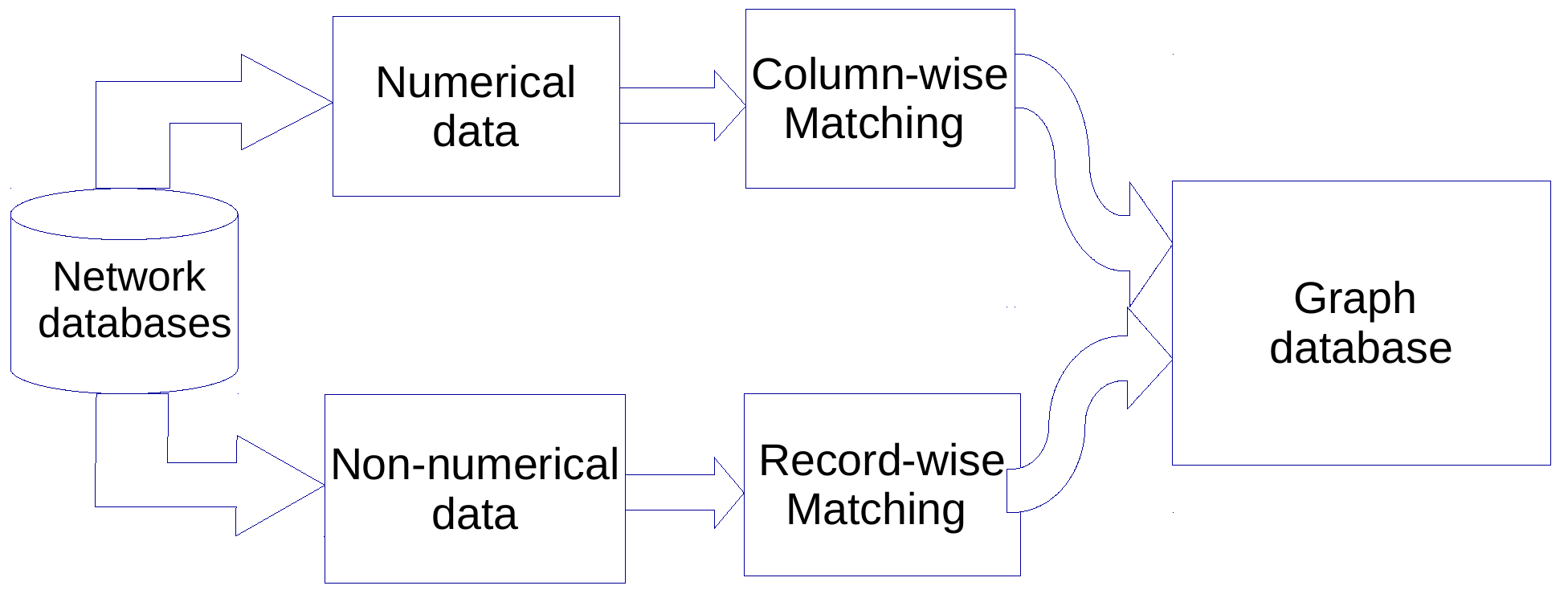}
			\caption{System diagram of automatic integration for network databases}
			\label{fig:overview}
		\end{figure}

		The system overview is shown in Figure \ref{fig:overview}. We divide the structured data into numerical data with only numerical fields and non-numerical data with only non-numerical fields. In each part, we develop an independent matching algorithm for field matching. Matching algorithms for numerical and non-numerical data are quite different, which will be introduced in section \ref{algNumericalMatching} and \ref{algNonNumericalMatching}. The results of each part are combined together to load into a graph database.
		
		\vspace{-0.2cm}
		
		\vspace{-0.3cm}
		\subsection{Numerical Matching}
		\label{algNumericalMatching}
		Numerical fields are table fields with records which are numerical values. For example, the $incident\_key$ and $prod\_key$ in Table \ref{tbl-ti} are numerical fields. Their record values construct a basis for similarity metrics of fields. We define each numerical field record values as a set. This is transfered to a problem of set similarity.
		
\begin{figure}[!th]
			\centering
			\includegraphics[width=3.2in,height=1in]{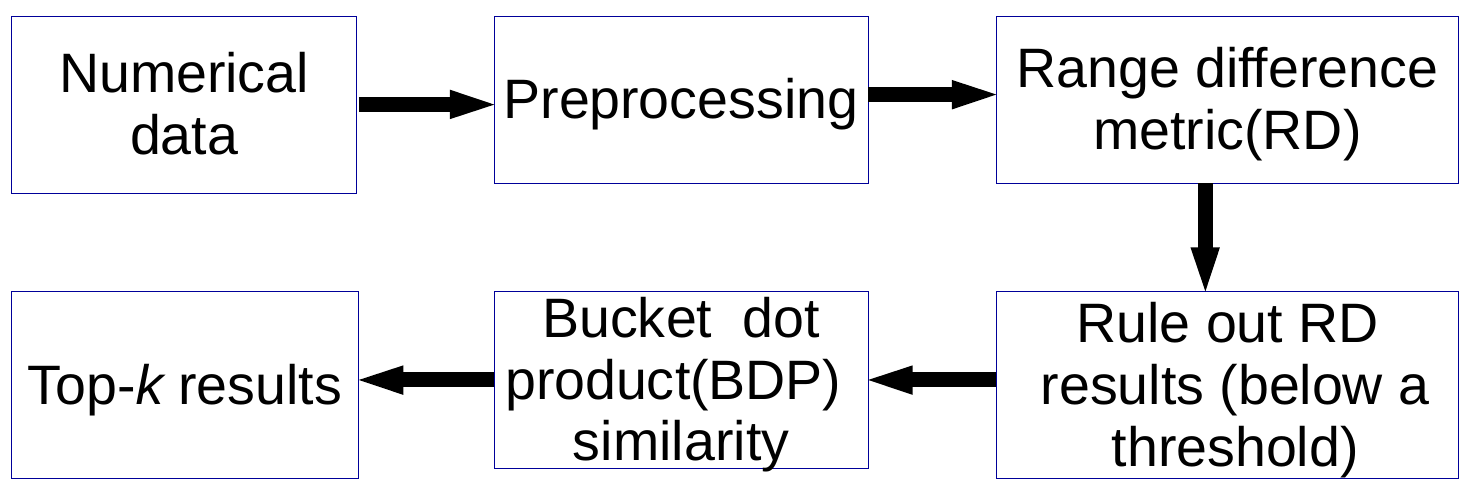}
			\caption{Numerical matching flow}
			\label{fig:numericalMatchingFLows}
		\end{figure}

		There are some common methods for solving set similarity including Jaccard index, Dice index, Hamming distance, cosine similarity \cite{cheatham2013string}, etc. However, it is not practical to just use one method to get accurate decision bounds of matching because of the noisiness and complexity of the structured databases. We propose a synthetic column-wise numerical matching algorithm to get the decision bounds to determine whether two fields are matched or not. The numerical matching algorithm is shown in Figure \ref{fig:numericalMatchingFLows}.
		
		\vspace{-0.2cm}
		
		The process of this algorithm is shown as follows:
		\begin{itemize} 
			\item  We preprocess the numerical data, such as removing null values, negative values and some exceptional non-numerical values in every field. 
			\item We apply range difference similarity metric to all the preprocessed numerical field pairs between every two tables.
			\item After we have got the range difference similarity score for each field pair, a threshold $T_r$ (which will be discussed later) will be decided to cut the filtered results. 
			\item Finally, bucket dot product similarity metric will be applied to the range difference similarity metric's filtered results. Then we sort the similarity scores of all the pairs to select the most correlated field pairs, that is, top-$k$ result as matched pairs with scores above a certain threshold $T_b$ will be selected to input to a graph database.
		\end{itemize}
		
		\subsubsection{Preprocessing}
		To deal with the irregular and noisy data, we do some preprocessing of field record values before the formal similarity calculations. This includes removing null values, negative values, any non-numerical values and considering unique values only. We do not consider negative values because they are useless and noisy data in the real databases. Almost all the table fields are about identifications, keys or numbers which can possibly be matched among them. In our databases, an average of $6\%$ of field record values are removed (not including unique value reduction) which does not impact the evaluation of instance-based algorithms.
		
		\subsubsection{Range Difference Similarity Metric}
		Considering the noisy and sparse characteristics of data, Jaccard similarity coefficient as a similarity metric, which measures how many common values between two sets, is not ideal for differentiating some matched pairs and non-matched pairs. For example, for an non-matched field pair with limited number of records, the number of common values in these two fields might take a large portion and hence the Jaccard similarity is very high for them. Their distribution of range, however, can be quite different, which is probably not to be matched in most cases. Therefore, we propose range difference (RD) similarity metric to measure the distribution of these field pairs first. Using RD similarity metric first, we can effectively and efficiently prune lots of unwanted computations, which can also hugely reduce time consumption for further matching. Given a field set $A$, we sort the record value and then get the different percentile (10th, 20th, 30th,..., 90th percentile). The percentile $ith$ value is recorded as $Ai$. Therefore, given two field record sets, $A$ and $B$, the RD value $D_i$ for each percentile $i$ is given as
		\begin{equation}
		D_i =  \frac{\left |A_i-B_i \right |}{|A_i+B_i|}
		\end{equation}
		We use 20th plus 30th percentile as the low range coverage, and 80th plus 90th percentile as the high range coverage, to cover the distribution of field record values. Hence the RD similarity score for a field pair (A, B) is defined as: 
		\begin{equation}
		RDS(A,B) = 1-\frac{D_{20}+D_{30}+D_{80}+D_{90}}{4}
		\end{equation}
		
		To keep consistent with the general similarity metric and be convenient for comparisons, we use 1 minus the averaged RD value as the RD similarity score RDS. The metric based on this similarity score is called RD similarity metric. Using this similarity metric, we can get the similar distribution for matched pairs. The value of $RDS$ is in [0, 1]. The bigger the similarity score, the more correlated the pair. There are several cases about the similarity score considered here: 
		
		(1) If there are no overlaps between two field ranges, $RDS$ would be as low as the minimum value 0.
		
		(2) If two fields have similar distributions, $RDS$ would be higher, up to 1.
		
		(3) If two field ranges overlap at the head, tail or in the middle, $RDS$ can fall into a middle value.
		
		In our toy example, the matched pairs $(incident\_key, incident\_id)$ and $(product\_id, prod\_key)$ have $RDS$ values as high as 0.996 and 0.999 respectively. In contrast, the pairs $(incident\_key, product\_id)$ and $(prod\_key, incident\_id)$ have no overlaps with $RDS$ value 0, which are not matched pairs. The more correlated the field pairs are, the higher RD similarity score they have. Therefore, using a threshold $T_r$ to filter results, we can almost rule out case (1) and part of case (3), then mainly consider case (2) to differentiate them further. To minimize the error of RD similarity score in the first step, we can use a conservative threshold close to the boundary to only filter out definite non-matched pairs, which will be discussed in the section \ref{ExperimentalEvaluation}.
		
		\subsubsection{Bucket Dot Product Similarity Metric}
		After we consider the distribution of field pairs with range difference  similarity metric, we propose bucket dot product (BDP) similarity metric  to further refine the filtered results of RD similarity metric. BDP similarity metric is to divide the whole concatenated ranges of two fields into different bucket/bins and compress each bucket as one point to calculate dot product similarity. The intuition behind this is that matched pairs generally have more common values than non-matched pairs. If we increase the bucket size up to a certain value to calculate dot product, it can make the similarities of all the non-matched pairs decrease more, and meanwhile make the similarities of all the matched pairs drop less, therefore it effectively increases the similarity gaps between matched pairs and non-matched pairs. Therefore, a good design of BDP will help significantly differentiate between matched pairs and non-matched pairs.
		
		The general dot product similarity of two vectors $X$ and $Y$ with $n$ elements is defined as follows: 
		\begin{equation}
		DP(X,Y) = \sum_{i=1}^{n}X_i\cdot Y_i
		\end{equation}
		
		We use the bucket number ($b_n$) to determine the number of buckets for calculating the dot product.
		Given two field record sets $A$ and $B$, we first derive the required vectors $A_v$ and $B_v$  for the input to the BDP similarity calculation. The vectors $A_v$ and $B_v$ derived from $A$ and $B$ are constructed in this way. Given two sets $A$ and $B$, we concatenate $A$ and $B$'s value ranges as a combined set $C$, and then divide $C$ into several buckets according to the $b_n$. If there is any one value in $A$ or $B$ falling in a bucket, the bucket point for $A_v$ or $B_v$' is 1, otherwise it is 0. Then we apply the general dot product similarity to $A_v$ and $B_v$. Therefore, the BDP similarity score (normalized) is defined as follows:
		\begin{equation}
		BDPS(A_v,B_v) = \frac{\sum_{i=1}^{b_n}A_{vi}B_{vi}}{|A_v||B_v|}
		\end{equation}
		%\justify
		where $b_n$ decides the sparsity/density of range distributions. Since sets $A$ and $B$ usually have different sizes with different ranges, it would make sense for $b_n$ the same for each set. For example, we calculate the BDP similarity  for a field pair $(incident\_key, incident\_id)$ in Table \ref{tbl-ti} and \ref{tbl-to} as $A$ and $B$ with $b_n = 3$. We first concatenate these two field ranges into a set \{201, 202, 203, 204, 207, 208, 209\}. Then we construct a set $C$ =\{\{201, 202, 203\},\{204, 207, 208\}, \{209\}\} according to the bucket number $b_n$. After that, we get the vector $A_v$ = \{1, 1, 1\} and $B_v$ = \{1, 1, 0\}. Finally, the BDP similarity score is 0.816, which is high for matching. If we set $b_n$ as 4, BDP similarity score for this pair is 1, which is the highest for matching.
		
	    $b_n$ is also an important factor to affect the quality of this metrics. According to our experimental observations, it is affected by the data range and distribution. Generally, matched pairs would have more similar ranges than non-matched pairs. A trade-off value of $b_n$ would effectively improve matched pairs' similarities more and also do not help improve non-matched pairs' similarities much, which can potentially increase more gaps between matched and non-matched pairs. The selection of $b_n$ will be discussed later in the section \ref{ExperimentalEvaluation}.
		
		\subsubsection{Primary Key Constraint Matching}
		If we consider all numerical fields and then do match computations between each other, it would be very time-consuming and not scalable when we have a large number of fields. If there are $n$ tables from database sources, and an average of $m$ fields in each table, the maximum $\binom{n}{2} * m^2$ comparisons are required. If we consider only table $A$ and table $B$ matching in a semantic way, the primary key constraint \cite{bhattacharjee2009ontomatch,  ioannou2019embench++} can be utilized to reduce the time for pair comparisons. Moreover, relational databases generally have primary keys and closely-related foreign keys in a table, which describe the semantic meanings of the table. It is important to match two tables in this semantic way to construct an effective graph database. Assuming every table has at least a primary key, we can identify the primary key and use the primary key's records to compare with all other keys records in other tables. We identify the primary key as a field that has unique record values before we apply RD and BDP similarity metrics. For example, $product\_id$ and $incident\_key$ are identified as primary keys in these two tables \ref{tbl-tp} and \ref{tbl-ti}, so we only need to compare $(product\_id, incident\_key)$, $(product\_id, prod\_key$, $(incident\_key, incident\_id)$ and $(product\_id, incident\_id)$ in total, reducing 33\% times of comparisons.
		
		\subsubsection{Final Top-$k$ Selection}
		To construct a high quality graph database, more true positive field pairs are preferable from higher similarity scores. Also, manual thresholds could lead to selection instability of field pairs with low similarity scores, so we seek top-k to further refine the quality of matching for the graph database. We sort all the candidate pairs by the similarity scores in a non-ascending order, then we verify this final field pair matching results to select top-$k$ field pairs, which involves only a little human labor.
		
		\vspace{-0.05cm}
		\subsection{Non-numerical Matching based on Top Priority Matching}
		\label{algNonNumericalMatching}
		We propose fast algorithms for non-numerical field matching--top priority match metric (TPM) for fast filtering, and match ratio score for final similarity computations. The diagram is shown in Figure \ref{fig:Non_numericalMatchingFLows}. The main process of our non-numerical matching algorithms is as follows:
		%The system flow is shown in Figure \ref{fig:Non_numericalMatchingFLows}.
		
		\vspace{-0.05cm}
		\begin{figure}[!th]
			\centering
			\includegraphics[width=3.1in,height=1.2in]{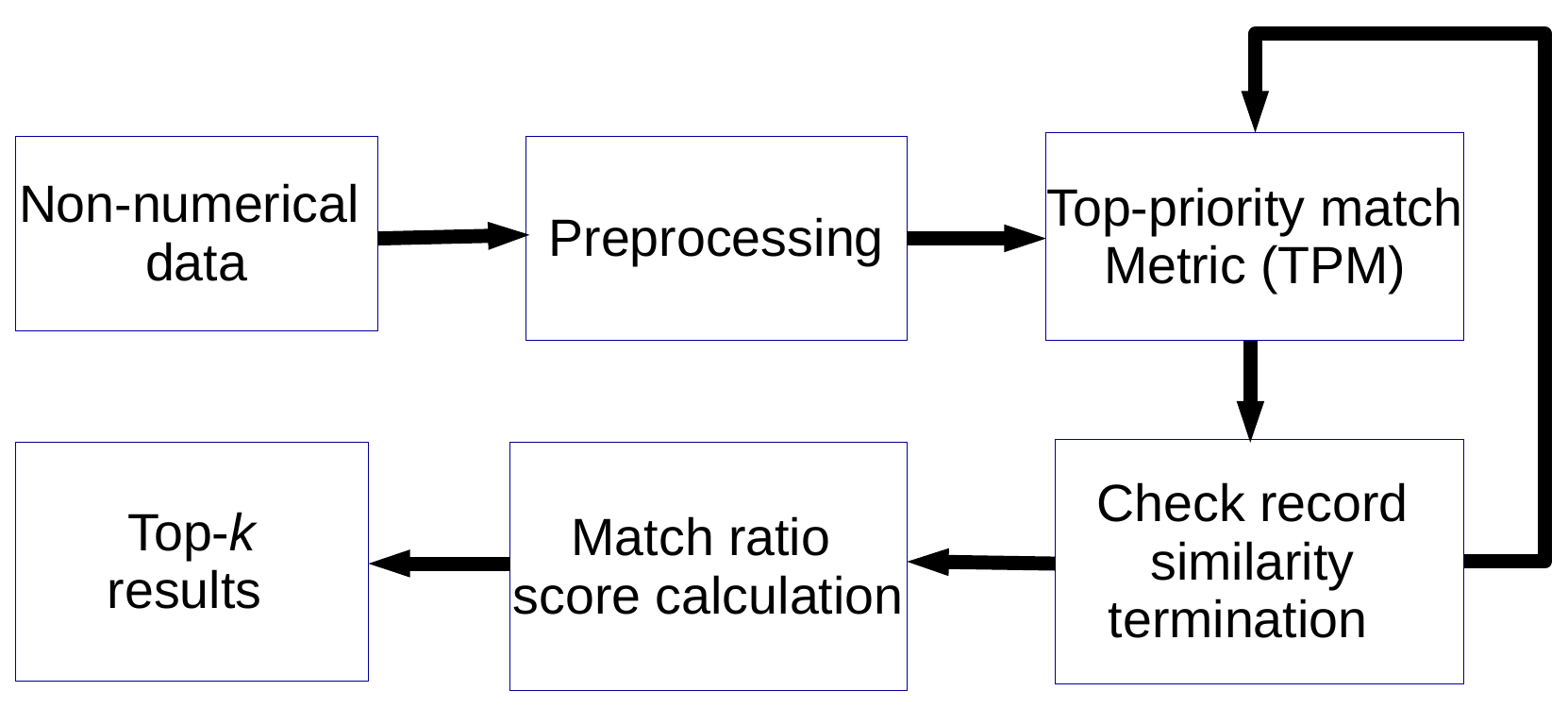}
			\caption{Non-numerical matching flow based on TPM}
			\label{fig:Non_numericalMatchingFLows}
		\end{figure}

		\begin{itemize} 
			\item Preprocess non-numerical data: this is an important step that decides the quality of our non-numerical matching algorithm. After splitting non-numerical data from the original databases, we use our designed natural language processing methods of segmentation, stemming and prefixing for every field record value.
			\item Calculate the record-wise similarities iteratively: 
			It is very time-consuming to apply cosine similarity to the combination of every record pair in a field pair. Considering the scalability of large-scale data matching, we propose top priority match metric for record-wise similarity calculation. In the process of iterative computations, we check the termination condition to terminate the iterations earlier, which significantly reduces time complexity.
			\item Calculate match ratio score: after record-wise similarity computations for every field pair are finished, we calculate the defined matching ratio score for each field pair.
			
			\item Select top-$k$ results: we sort the field pairs by the matching ratio score in a non-ascending order, and the top-$k$ field pairs are selected as the final results as an input into a graph database. 
			
		\end{itemize}
		Here, we discuss each proposed steps in detail.
		\subsubsection{Preprocessing}
		\label{PreprocessingNonNumerical}
		Non-numerical matching considers partial match between two strings. For example, product \dblquote{cisco0510} and product \dblquote{cisco0500} are in the same series, which is considered as a partial match. Hence, we propose the following preprocessing method. (1) Parse every record string $A$,  remove null value, separate alphabetic and numerical characters into different new substrings, and tokenize the string words. (2) Stem the alphabetic strings of the original record and new substrings. (3) Reserve the prefixes with a certain length of the original numerical strings and the new substrings if they are digital substrings. The prefix length is 2 here according to the general characteristics of network databases and our experiments. 
		
		Each record string is preprocessed in those three steps above. For example, we have an original field record $R$ \{'mem-4700m-64d='\} in Table \ref{tbl-to}. We can obtain a string collection $X$ \{'4700', '64', '4700m', 'd', 'm', 'mem 4700m 64d', '64d', 'mem', '47xx'\} after preprocessing $R$.
		
		\subsubsection{Top Priority Match Metric for Record-wise Similarity}
		One intuitive way is to preprocess all the combination of record pair comparisons and calculate the similarity of each record-wise pairs. That would be very time-consuming or even unfeasible when the data are large. Specially, if two fields are not correlated as a matched pair, it would be costly for useless computations. Therefore, we propose a fast record-wise matching algorithm called top priority match metric for record-wise similarity (TPM) to fulfill this. Intuitively, if two fields $A$ and $B$ are correlated, there will be a high percentage of record-wise pairs that have higher similarities. The probability of a matched record pair encountered is higher than non-matched record pairs. Therefore, we first sort all the preprocessed records in each two fields $A$ and $B$, then we compute how many of records in $A$ are matched with records in $B$ from top to bottom, and vice-versa.
		The comparisons can hence be terminated as long as the current record pair similarity achieves below the similarity threshold $T_{rn}$ we set for deciding the matching of a record pair, which greatly reduce the times of comparisons with combinations.
		
		%That is applied to our proposed matching ratio score, which will be discussed in the next section. 
		
		\subsubsection{Record Pair Similarity}
		In this fast record-wise comparisons, the record pair similarity used is cosine similarity between two record collections after preprocessing two records. It decides how and when to reduce the comparisons of matching in a fast and effective way. A threshold $T_{rn}$ is to decide how similar a record pair is as a matched record pair, which can also be adjusted by users. 
		
		Given a preprocessed string collection $X$ and another preprocessed string collection $Y$, we remove duplicated elements and transfer them into a set $XY$($X$ union $Y$). Next we generate a binary vector $V_x$ and a binary vector $V_y$ according to the value distribution of $X$ and $Y$ in $XY$, then we calculate the cosine similarity $sim(V_x,V_y)$ between $V_x$ and $V_y$ by getting their dot product divided by their magnitude multiplication.
		\begin{equation}
		sim(V_x,V_y) = \frac{V_x\cdot V_y}{\left | V_x \right |\left | V_y \right |} 
		\end{equation}
		For example, we have a preprocessed string collection $X$ \{cisco, 0510, cisco0510, 05xx\}, and $Y$ \{cisco, 05xx, 0500, cisco0500\}, we transfer them into a set of $X$ union $Y$, $XY$ \{cisco, 0510, cisco0510, 05xx, 0500, cisco0500\}. Then the binary vectors generated according to $X$, $Y$ and $XY$ are $V_x$ \{1, 1, 1, 1, 0, 0\} and $V_y$ \{1, 0, 0, 1, 1, 1\}. Finally, we calculate the cosine similarity of $V_x$ and $V_y$ as the similarity of $X$ and $Y$, that is, $sim(X,Y)= (1+1)/(2*2) = 0.5$.
		
		\subsubsection{Matching Ratio Score}
		Matching ratio score is proposed to calculate the final similarity for a field pair. After we have gone through the reducing comparisons for record-wise similarities, we select the number of record pairs that have similarity scores above $T_{rn}$. A matching ratio score as a final field pair similarity is the average of ratios of top matched record pairs calculated as follows:
		%to select non-numerical candidate field pairs 
		
		Given two non-numerical sets $A$ and $B$, there are $m$ items $\left \{a_1, a_2,...,a_m \right \}$ in $A$ and $n$ items $\left \{b_1, b_2,...,b_n \right \}$ in B. The matching ratio score(MR) between $A$ and $B$ is defined as follows.
		
		\begin{equation}
		MR(A, B) = \frac{1}{2}*(\frac{\sum_{i=1}^{m} A_i}{m} + \frac{\sum_{j=1}^{n}B_j}{n})
		\end{equation}
		where 
		\begin{equation}
		A_i = \begin{cases}
		1 & \text{ if } \exists \hspace*{0.11cm} b_j \in B, \hspace*{0.1cm}  {sim(a_i, b_j) >= T_{rn}}  \\ 
		0 & \text{ otherwise }  
		\end{cases}
		\end{equation}
		and 
		
		\begin{equation}
		B_j = \begin{cases}
		1 & \text{ if } \exists \hspace*{0.11cm} a_i \in A, \hspace*{0.1cm}  {sim(b_j, a_i) >= T_{rn}}  \\ 
		0 & \text{ otherwise }  
		\end{cases}
		\end{equation}
		where $sim(a_i, b_j)$ and $sim(b_j, a_i)$ are the cosine similarities of the record pairs $(a_i, b_j)$ and $(b_j, a_i)$, respectively. $MR$ value is in [0, 1] and it is the final similarity score to decide the correlation of each field pair.
		
		\subsubsection{Final Top-$k$ Results}
	     Similar to numerical field matching, matched non-numerical field pairs in the result list are more meaningful and important than non-matched field pairs, so we select top-$k$ results of non-numerical field pairs sorted with matching ratio scores in non-ascending order as an input to a graph database. $K$ value can be selected by users for deciding most effective field pairs in a graph database and limiting the size of the graph database.
		\vspace{-0.05cm}
		\subsection{More Efficient Hashing Algorithm for Non-numerical Matching}
		The proposed TPM metric can be effective to distinguish between matched and non-matched non-numerical fields. However, it possibly involves all the pairwise record combinations in the worst time complexity, which is time-consuming for large databases with millions of records. Therefore, we propose applying more scalable minHash-locality sensitive hashing algorithm (MH-LSH) \cite{leskovec2014mining} to estimate the matching score of non-numerical fields in the databases. It can greatly reduce the comparison size and time for non-numerical record-wise pairs with little cost of matching accuracy down.
		
\begin{figure}[!th]
			\centering
			\includegraphics[width=3.1in,height=1.1in]{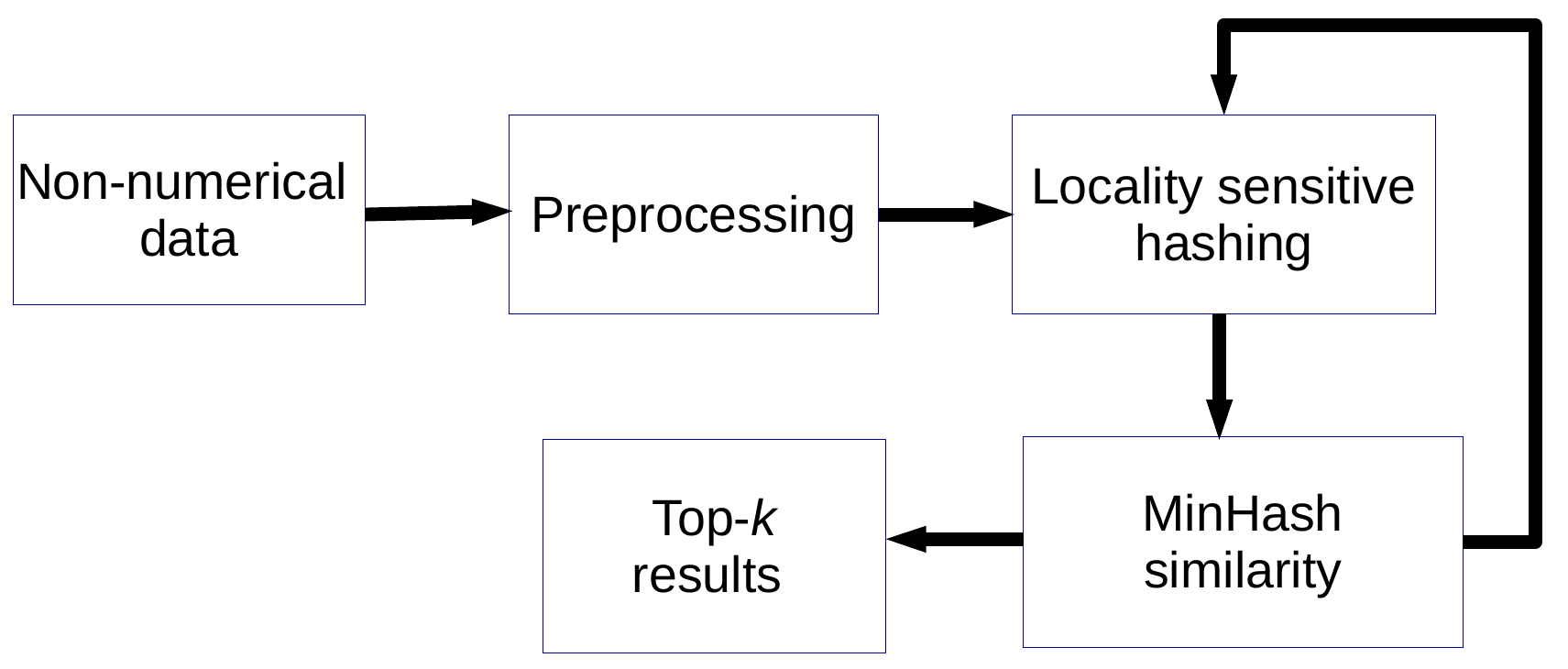}
			\caption{Non-numerical matching flow based on MH-LSH}
			\label{fig:Non_numericalMatchingFLowsHash}
		\end{figure}

		The proposed diagram for non-numerical matching based on MH-LSH is shown in Figure \ref{fig:Non_numericalMatchingFLowsHash}. The main process of the algorithm is as follows:
		\begin{itemize} 
			\item Preprocess non-numerical data: this step is the same as the preprocessing step of TPM algorithm.
			\item Select matched field pairs fast: we apply the locality sensitive hashing technique in the database field matching for fast selecting field pairs that are correlated. In this process, it uses minHash similarity to decide the filtering results for field pairs.
			\item Field pair similarity calculation: We use minHash technique in the databases to estimate the matching score of field pairs.
			\item  Select top-$k$ results: Same as the operation for non-numerical matching based on TPM, we sort the field pairs by the estimated matching score in a non-ascending order, the top-$k$ field pairs are selected as the final results of field pairs as an input to a graph database.
		\end{itemize}
		
		\vspace{-0.2cm}
		Here, we discuss important steps in detail.
		
		\subsubsection{Matching Score Estimation with MinHash}
		The similarity score of matched field pairs can be estimated fast with minHash signatures.
		Given two fields $A$ and $B$, we can evaluate the similarity between them as follows. We choose $n$ hash functions $h_1, h_2,..., h_n$. For each hash function $h_j$, we let a signature of set $A$ be $sgn({A})=\min_{i:a_{i}\in {A}} h_j(a_i)$ for $j \in n$. Let a signature of set $B$ be $sgn({B})=\min_{i:b_{i}\in {B}} h_j(b_i)$ for $j \in n$. Then, the probability that the two sets have the same minHash signatures can be used to estimate the similarity between them.  
		\begin{equation}
		{\rm MHSim}({A},{B})=P\{sgn({A})=sgn({B})\}
		\end{equation}
		
		Here each record in a field $A$ or $B$ is also preprocessed with the same preprocessing method of TPM algorithm.  The preprocessed record strings are combined into a new set like a word set in a document. We also use $k$ shingle (a substring of length $k$) to create a set of $k$-shingles strings and apply $n$ different hash functions on the set of strings.

		\subsubsection{Field Pair Selection with Locality Sensitive Hashing}
		Performing pairwise similarity measurement can be time consuming with large amounts of field pairs available. In order to identify which field pairs are similar quickly, we propose using locality sensitive hashing (LSH) to select the candidate field pairs. 
		
		The values of minHash signature $sgn(A)$ for one field $A$ are grouped into $b$-tuples (referred to as sketches) with $r$ rows. Similar field pairs have similar minHash signatures and hence have a high probability of having the same sketches. Moreover, dissimilar pairs have low chance of falling into the same sketch. The probability that two fields of $A$ and $B$ have at least one sketch (of size $b$) in common out of $r$ is
		\vspace*{-2pt}
		\begin{equation}
		P_{C}({A}, {B})=1-(1- {\rm MHSim}({A},{B})^{b})^{r}
		\end{equation}
		
		Therefore, we can find the candidate pairs with the designed number $b$ and $r$. The selection of $b$ and $r$ is generally decided by a threshold $t = (1/b)^{1/r}$ shown in \cite{leskovec2014mining}, which indicates how similar the two fields is to be considered as a candidate pair, and can also be set by users. In this way, if pairs with similarity above $P_{C}({A}, {B})$ , they will be selected as candidate pairs to be further estimated, and the matching score between them with minHash will be calculated.
		
	   \section{Classification-based Matching}
		\label{sec_classification}
		Our proposed algorithms for numerical and non-numerical matching involve several manual thresholds to get the matching results. The manual threshold selection is trivial and may not be generic to new databases. We propose a classification-based method to decide the matching or non-matching.
		Using our proposed similarity scores as features, we construct a classification-based matching system.

	    For a classification method, there are four important factors involved--input data, feature design, model selection and training/testing. The details of classification are introduced in the following sections.
	   \vspace{-0.1cm}
	     \subsection{Sampling for ground truth}
	    To address the problem of limited availability of the ground truth in our Cisco dataset, we use a sampling method \cite{kohler2010sampling} to synthesize more ground truth pairs, which preserves the "synchronization property" (to preserve the Jaccard similarity of original sets).
	    Given an original ground truth $A$ and $B$, we sample a new ground truth pair based on $A$ and $B$. The sampling process is to make sure if a particular sample $e$ is sampled in $A$ and $e$ is also in $B$, then it is also sampled from $B$ \cite{kohler2010sampling}.
	    Assume we have only $X$ real numerical ground truth pairs that are real data from the databases and labeled by humans.
	    To extend ground truth pairs, we synthesize 100 times of ground truth pairs from each of the field pairs.  For example, in our dataset, there are 60 real numerical ground truth pairs that are real data from the databases and labeled by humans, so we  synthesize  6000  more ground  truth  pairs  which  are  enough  to  apply  to  a  classifier. Synthetic non-numerical ground truth pairs are also generated in the similar way.
	    
	   \vspace{-0.1cm}
	    \subsection{Feature Design for Classification}
	    In the previous numerical field matching, we have generated range difference (RD) similarity and bucket dot product (BDP) similarity metrics. We propose to use these similarity scores as features. There is one important manual threshold-bucket number involved while calculating BDP similarity, so we generate 15 different BDP similarity scores based on different bucket numbers in [100, 200, 500, 1000, 2000, 5000, 10000, 50000, 100000, 500000, 1000000, 5000000, 10000000, 50000000, 100000000], which are obtained from the data range of our real ground truth. As a consequence, for each filed pair instance, we have 16 features of scores in total.

        For non-numerical field pairs, we have proposed matching ratio score and minHash similarity metric to decide the similarity of these pairs. During the calculation, we have one important record-pair threshold $T_{rn}$ while calculating matching ratio score. Therefore, we use 4 different $T_{rn}$ values  in [0.3, 0.4, 0.5, 0.6] to calculate different matching ratios as features.
        In total, 5 features of scores are obtained for each non-numerical field pair.
        
		\vspace{-0.2cm}
        \subsection{Training and Testing with SVM }
        We select Support Vector Machine (SVM) as our classifier for the reason that SVM works well on unstructured data such as non-numerical text and scales well with high dimensional data.
		We split our data into $90\%$ as training and validation data, $10\%$ as the test data. In training stage, 5-fold cross validation is used for training and validation, and the results are shown in the experiment.  % blue color

		\section{Experimental Evaluation}
		\label{ExperimentalEvaluation}
		In the section, we evaluate our technique NEMA for structured network database matching. To be specific, we measure the effectiveness of NEMA using ground truths for numerical and non-numerical data that are annotated by humans. Meanwhile, experiments on a large dataset are also conducted, and we show the top-k effective results of matching field pairs. Moreover, comparisons of NEMA with other baseline methods are also shown.
	   \vspace{-0.1cm}
		\subsection{Data}
		The structured network management databases available in the form of database tables are provided by Cisco Systems, Inc. The dataset includes \dblquote{install\_base} and \dblquote{service\_request} databases. They are generated by different departments and groups of the corporation, which are heterogeneous and diversely distributed.
		
		In these databases, there are 21 tables which contain 1,458 columns. Each column has 10 million records on the average. Out of them, there are 679 numerical fields and 779 non-numerical fields. Therefore, a complete match involves the maximum 1,067,882 field matching decisions. With primary key constrains in numerical matching, there are 5 \dblquote{primary keys} on average in each table, which would reduce to 374,326 field pairs matching.
		
		We have ground truth field pairs that are annotated by humans to be matched or non-matched field pairs for a subset of the data. There are 60 field pairs of ground truth in numerical data and 40 pairs in non-numerical data. For future reference, table names in service\_request database start with \dblquote{T\_}, and start with \dblquote{X\_} in install\_base database, respectively. 
		
	   \vspace{-0.1cm}
		\subsection{Experimental Setup}
		We implement NEMA system in Python. To evaluate the effectiveness of NEMA , we evaluate the numerical and non-numerical algorithm parts, respectively. For each part, we first evaluate its algorithms based on the ground truth data. Then all the column pairs in the large dataset are evaluated in the following experiments shown in sections \ref{exper-numericalTOpResult} and \ref{exper-nonnumericalTOpResult}, which shows the effectiveness of NEMA. Finally, we compare with the common matching system COMA \cite{aumueller2005schema} on the ground truth for both schema-level and instance-level matching.
		
		To construct an effective graph database, identifying accurately matched pairs and reducing non-matched pairs appearing at the top are much more important than finding more non-matched pairs. Also, our evaluation is based on the balanced ground truth of positive and negative pairs. Therefore, we use an overall metric ``accuracy" ($ACC$) to evaluate our field matching algorithms. $ACC$ is calculated with true positive($TP$), true negative($TN$), total positive pairs($P$) and total negative pairs($N$) (Here positive means matched and negative means non-matched). $
		ACC = (TP + TN)/(P + N).$

		\subsection{Evaluation based on Numerical Data}
		We evaluate our technique NEMA on the numerical data in two parts. We use numerical ground truth to evaluate the effectiveness of NEMA numerical algorithm. Then the matching results of all of other numerical field pairs are also described.
		
		\subsubsection{Evaluating of Ground Truth}
		We show the evaluation result of NEMA numerical algorithms and the compared baseline method-Jaccard similarity using numerical ground truth (We choose Jaccard similarity since it is an exemplar method considering common values of two sets for similarity calculation). In the dataset, there are 30 matched ground truth field pairs which are originally from fields pairs annotated by humans or from join operations in the databases and proved to be matched field pairs. We randomly sample 300 non-matched field pairs confirmed by humans and select 30 non-matched field pairs from them as non-matched ground truth.
		
		\begin{table}[!th]
			\centering
			\caption{Example of numerical ground truth field pairs}
			\label{tbl:gt_pos_num_10}
			\scriptsize
			\begin{tabular}{p{0.3cm}p{2.4cm}p{2.5cm}p{1.5cm}}  %{@{}|cllc|@{}} 
				\toprule
				\textbf{No.} & \bf{Table.fieldA}  & \textbf{Table.fieldB}                       & \textbf{Matched class} \\ \midrule
				1   & T\_INCI.prod\_hw\_key        & T\_HW\_PROD.bl\_prod\_key         & 1              \\ 
				2   & T\_INCI.cur\_ct\_key  &
				T\_CT.bl\_ct\_key                     & 1              \\ 
				3   & T\_INCI.up\_tech\_key      & T\_TECH.bl\_tech\_key      & 1              \\ 
				4   & T\_INCI.inci\_id                 & T\_INCI\_I2.inci\_id               & 1              \\ 
				5   & T\_INCI.bl\_cot\_key             & T\_COT.bl\_cot\_key                  & 1              \\ 
				6   & T\_INCI.inci\_id                 & T\_OR\_HD.inci\_id                  & 1              \\ 
				7   & T\_INCI.item\_id          & T\_PROD.item\_id                    & 1              \\ 
				8   & T\_INCI.ins\_site\_key       & T\_SITE.bl\_site\_key                             & 1              \\ 
				9   & T\_OR\_LN.prod\_key                       & T\_PROD.bl\_prod\_key                       & 1              \\ 
				10  & T\_OR\_HD.header\_id                       & T\_OR\_LN.header\_id                      & 1              \\ 
				1   & T\_OR\_HD.order\_dur       & T\_OR\_LN.loc\_key & 0              \\ 
				2   & T\_CAL.bl\_cal\_key       & T\_PROD.item\_id                    & 0              \\ 
				3   & T\_INCI\_I2 & T\_DEFT.deft\_id                              & 0              \\ 
				4   & T\_INCI\_I2.res\_time       & T\_TECH.sub\_tech\_id                 & 0              \\ 
				5   & X\_PRO.list\_price             & T\_TECH.sub\_tech\_id                 & 0              \\ 
				6   & T\_INCI\_I2.res\_time       & T\_TECH.bl\_tech\_key      & 0              \\ 
				7   & T\_OR\_HD.deliv\_dur    & T\_DEFT.deft\_key                         & 0              \\ 
				8   & T\_OR\_LN.hold\_dur                   & T\_TECH.sub\_tech\_id                 & 0              \\ 
				9   & T\_IN.serlevel\_key      & T\_INCI.last\_dur        & 0              \\ 
				10  & T\_IN.closect\_key       & T\_INCI.resp\_tz    & 0              \\ \bottomrule
			\end{tabular}
		\end{table}
		
		Table \ref{tbl:gt_pos_num_10} shows 10 matched and 10 non-matched field pair examples of numerical ground truths. Columns ``Table.field A'' or ``Table.field B'' shows a table name and its field pair name to be compared. The matched class column indicates that the field pair is matched with value ``1'' or non-matched with value ``0''.  For example, in the first row, \dblquote{T\_INCI.prod\_hw\_key} indicates a field \dblquote{prod\_hw\_key} in the table  \dblquote{T\_INCI},  and \dblquote{T\_HW\_PROD.bl\_prod\_key} indicates a field \dblquote{bl\_prod\_key} in the table \dblquote{T\_HW\_PROD}. This field pair about products' key is matched, which is indicated with ``1'' in the matched class value. The remaining rows share the same characteristics as well.
		
		\vspace{-0.1cm}
		\begin{figure}[!th]
			\centering
			\includegraphics[width=3.2in,height=1.9in]{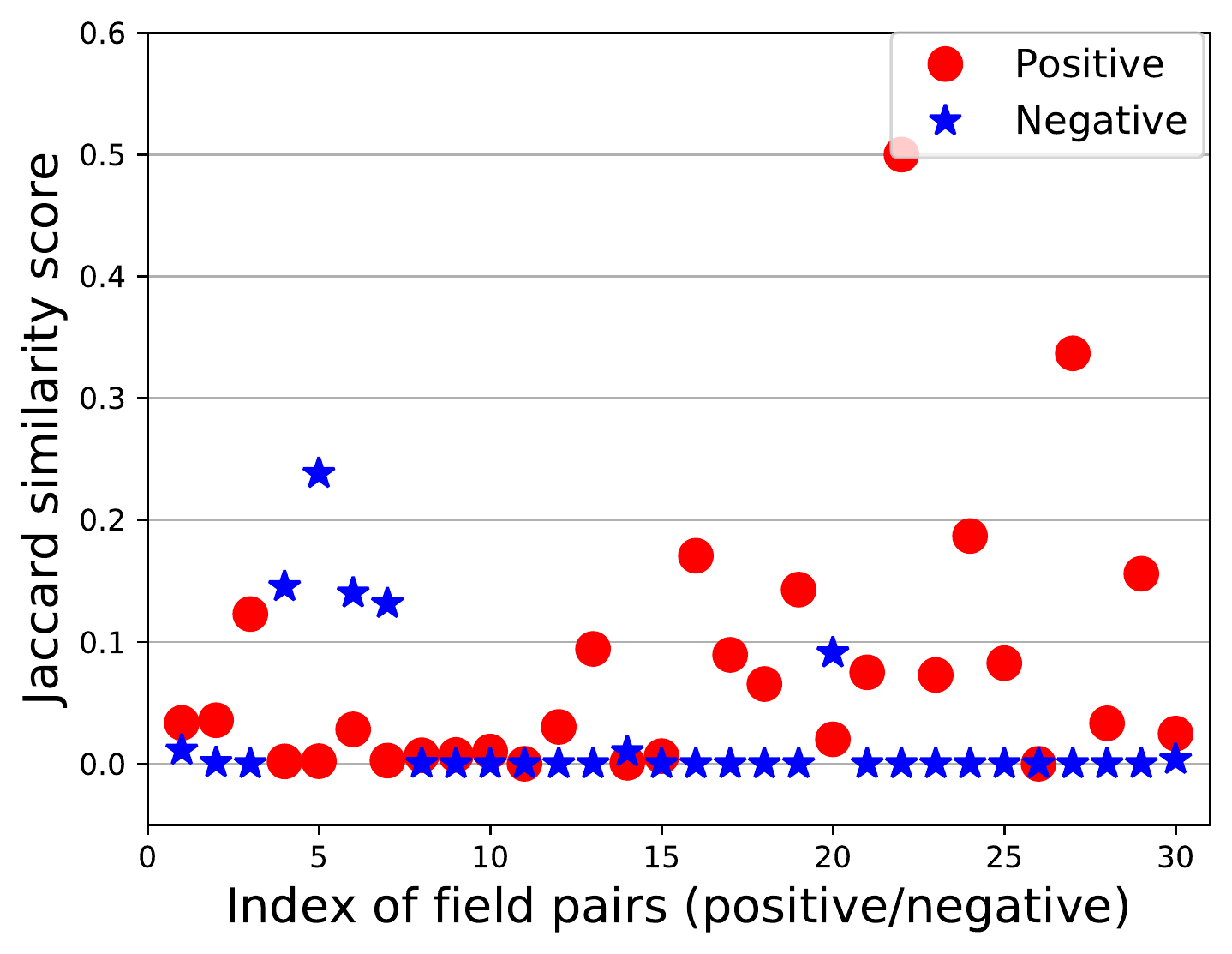}
			\caption{Jaccard similarity metric on the numerical ground  truths}
			\label{fig:paper_JaccardGT60}
		\end{figure}
		
		The problem of similarity of a numerical field pair is modeled as the problem of similarity of a set pair. The baseline method for the similarity of a numerical field pair is the well-known Jaccard similarity which measures the similarity of two given sets.
		
		%/home/fubao/workDir/ResearchProjects/CiscoWISH/CreateGraph/IdentifyExistingRelationship/byValue/evaluation/range_evaluation
		Figure \ref{fig:paper_JaccardGT60} shows the Jaccard similarity scores on these ground truth field pairs. Red circle represents matched pairs and blue star represents non-matched pairs. $X$ axis denotes the index of these 30 matched and 30 non-matched field pairs, and $Y$ axis indicates Jaccard similarity score.  From this figure, we can see that there are about half of positive and negative pairs mixed together from which are difficult to differentiate. 
		\vspace{-0.01cm}
		\begin{figure}[!th]
			\centering	\includegraphics[width=3.2in,height=1.9in]{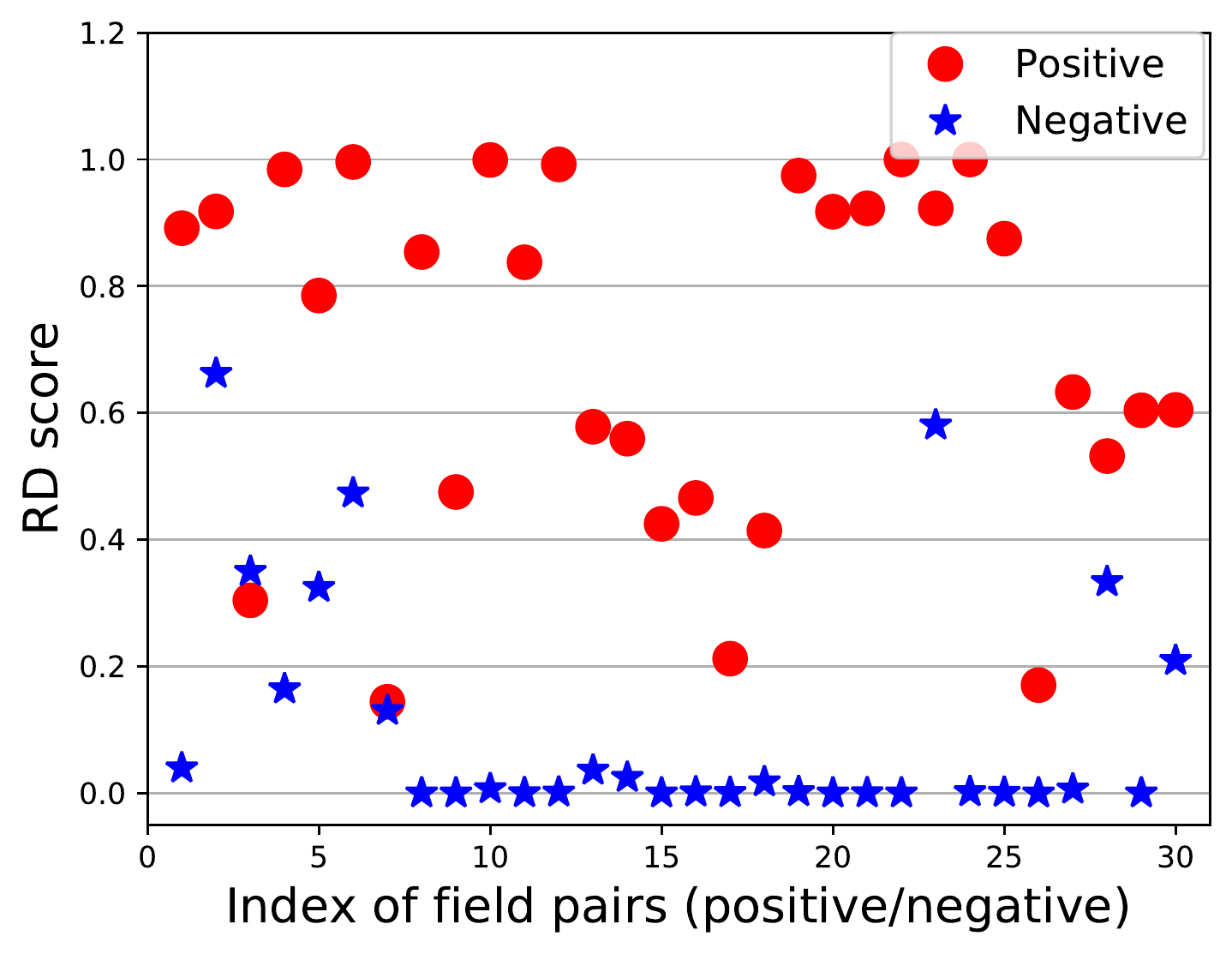}
			\caption{RD similarity metric on the numerical ground truths}
			\label{fig:paper_RangeDiff}
		\end{figure}
		
		Figure \ref{fig:paper_RangeDiff} shows the RD similarity metric on these ground truth field pairs. We know that a pair is more likely to match if its RD similarity score is larger. There are about only $30\%$ positive and negative pairs mixed together, which has greatly improved the result over the Jaccard similarity result.
		
		From the previous result of RD similarity metric, we can set a low threshold $T_r$ and then rule out the result pairs below $T_r$ as non-matched field pairs. To minimize the error of initial matching, we can select a conservative threshold $T_r$ close to 0.1 (all of the uncertain pairs are above 0.1) to eliminate some generally precise non-matched pairs.
		
		After we filter field pairs from RD similarity metric, there are totally 39 field pairs left, that is, about $35\%$ of all ground truth pairs (all of them are non-matched field pairs) are pruned. In these 39 pairs, there are 30 matched pairs and 9 non-matched pairs waiting to be input to the BDP similarity metric.
		
\begin{figure}[!th]
			\centering
			\includegraphics[width=3.2in,height=1.9in]{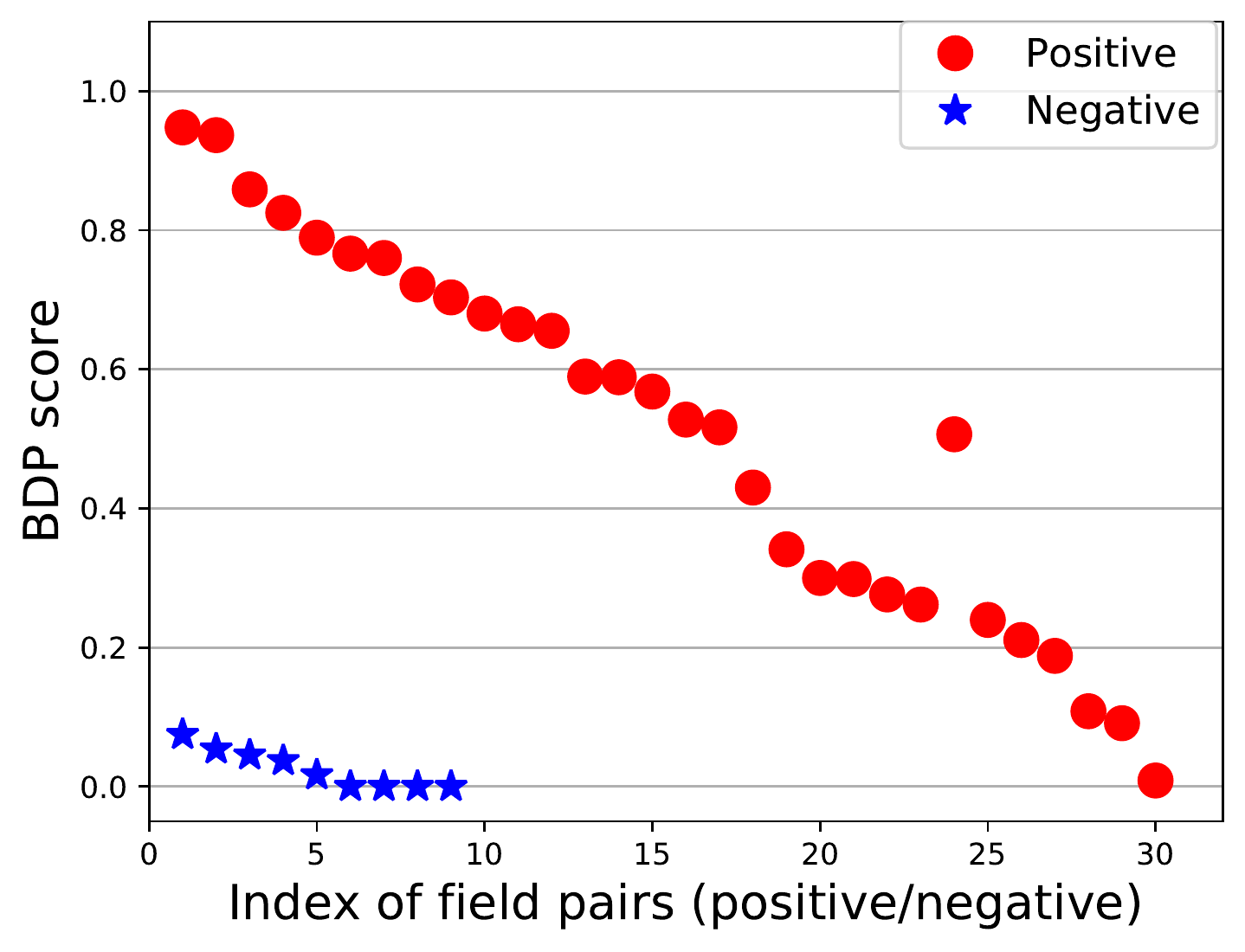}
			\caption{BDP similarity metric on the remaining numerical ground truth pairs }
			%after range difference similarity metric filtering
			\label{fig:paper_NumBucketDP34}
		\end{figure}

		Figure \ref{fig:paper_NumBucketDP34} shows the BDP result after the previous RD similarity metric result. $X$ axis denotes the index of these 30 matched and 9 non-matched pairs, and $Y$ axis indicates the normalized BDP similarity score in a non-ascending order. We can see from that the BDP result has very good decision line between matched pairs and non-matched pairs in which the final threshold $T_b$ is chosen around 0.1. The final accuracy can be up to 95\% when $T_b$ is 0.1. 
% ResearchProjects/CiscoWISH/CreateGraph/IdentifyExistingRelationship/byValue/evaluation/decisionTree_rangeScoreToDotProduct/paper_plot
		\vspace{-0.2cm}
		
\begin{figure}[!th]
			\includegraphics[width=3.2in,height=1.9in]{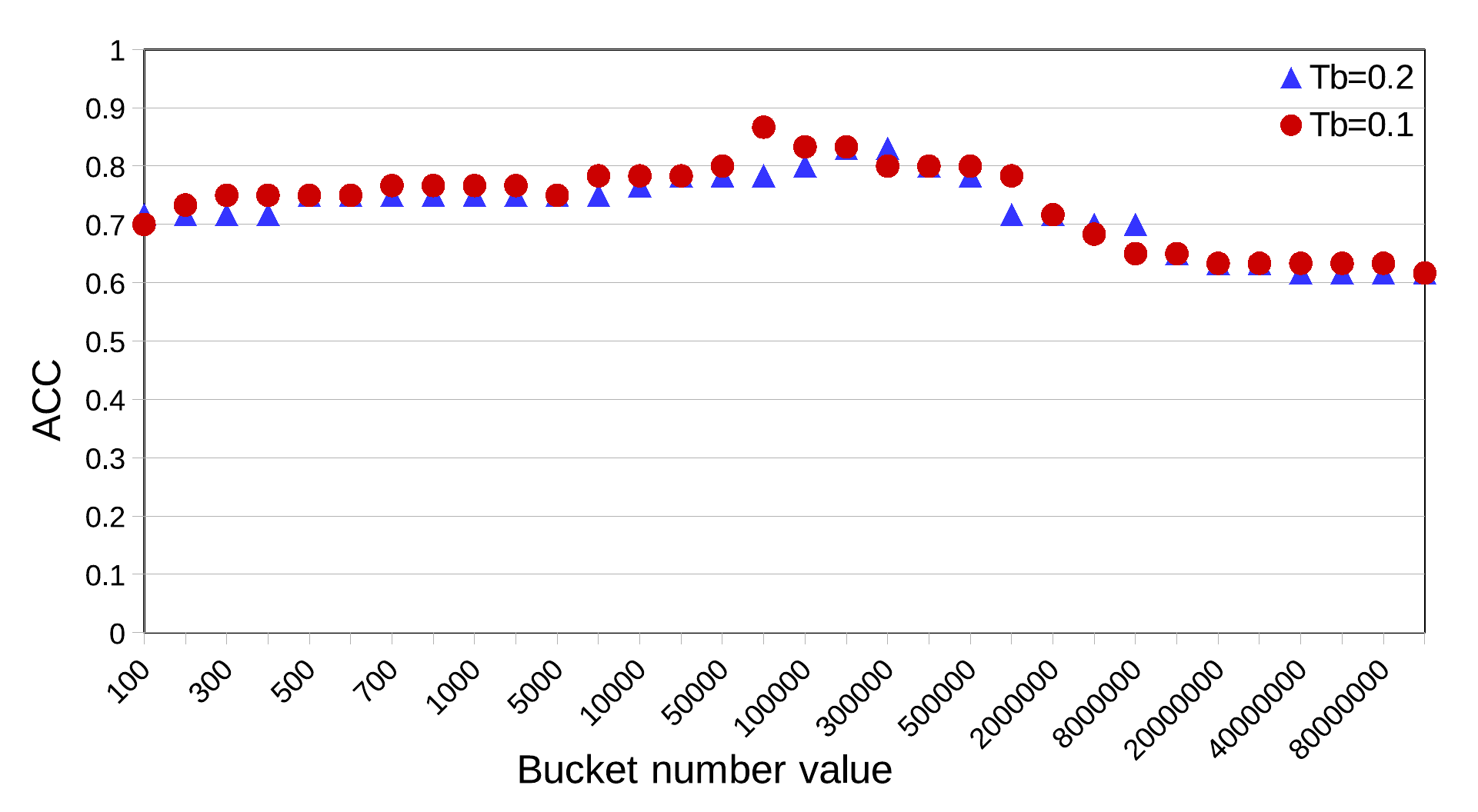}
			\caption{BDP accuracy comparison with different bucket numbers on the numerical ground truths }\label{fig:BucketNumTrendAnalysis}
		\end{figure}

		The bucket number $b_n$ is the main factor affecting the BDP similarity and the final results. To evaluate the effectiveness of different $b_n$ values, we use all the numerical ground truth to test the accuracy with different $b_n$ values from 100, 200, 300, 400, 500 up to 100,000,000 where there exists some numerical records that have values up to 10 billion in almost all the fields. Figure \ref{fig:BucketNumTrendAnalysis} shows the accuracy of BDP similarity metric with different $b_n$ values when the threshold $T_b$ is set at 0.1 and 0.2. It shows a similar summit that the accuracy is at an optimal value when the $b_n$ is around from 10,000 to 200,000. BDP accuracy goes down when  the $b_n$ becomes smaller or bigger. We use a trade-off value $b_n= 50,000$ in our experiments to avoid overfitting.
		
		%/home/fubao/workDir/ResearchProjects/CiscoWISH/CreateGraph/IdentifyExistingRelationship/byValue/evaluation/decisionTree_rangeScoreToDotProduct/paper_plot
		\vspace{-0.4cm}
			
		\subsubsection{Top-$20$ Similarity Results}
		\label{exper-numericalTOpResult}
		The matching experiment based on all the 679 numerical fields is shown here. Table \ref{tbl:top20NumericalReuslt} shows top-$20$ matching results. We use RD similarity score threshold $T_r = 0.1$ and bucket number $b_n = 50,000$ for BDP similarity computation. All the rows are accurate matches, which are also confirmed by humans. The accuracy can be up to 100\% for the top-$20$ results, which shows a great potential for our numerical matching algorithm applied on the large dataset, also reducing lots of human labor for matching. Moreover, with our system-aided matching findings, we can find some pairs matching which are difficult to be found with human annotations such as \dblquote{changewg\_key} and \dblquote{subregion\_key}, indicating which regions that the workgroup mainly serves. 
		%It achieves very high accuracy from top matching results and find a large portion of matching pairs which dramatically reduce the human labor and further facilitate for our graph query and device management, etc.
		
		To ensure a graph database more meaningful and complete, the selection of top-$k$ results is finally decided by users. Users can observe the top-$k$ results and rule out the unwanted matches as the final input pairs to a graph database so as to introduce less \dblquote{noisy} connections of the graph database. 
		
		% \usepackage{booktabs}
		\begin{table}[!th]
			\centering
			\caption{Top-$20$ similarity result of numerical field pairs}
			\label{tbl:top20NumericalReuslt}
			\scriptsize
			\begin{tabular}{p{2.8cm}p{2.9cm}p{1.8cm}}               %{@{}lll@{}}
				\toprule
				\bf{Table.field A}                                                                & \bf{Table.field B} & \bf{BDP similarity}                   \\ \midrule
				T\_DEFT.deft\_key          & T\_INCI\_DE.bl\_def\_key       & 0.844        \\
				T\_OR\_LN.item\_id            & X\_PRO.item\_id          & 0.698        \\
				T\_DEFT.deft\_id                              & T\_INCI\_DE.defect\_id            & 0.682         \\
				T\_INCI.item\_id       & X\_PRO.item\_id          & 0.673         \\
				T\_DEFT.deft\_key               & T\_PROD.item\_id   & 0.64 \\
				X\_INS.item\_id         & X\_PRO.item\_id          & 0.597         \\
				\textbf{T\_PROD.bl\_prod\_key}         & \textbf{T\_SUR.task\_key}       & \textbf{0.567} \\
				T\_INCI.changewg\_key & T\_WK.subregion\_key             & 0.551         \\
				T\_INCI.changewg\_key & T\_WK.wkgrp\_key           & 0.551          \\
				T\_INCI.changewg\_key & T\_WK.theater\_key         & 0.551        \\
				T\_INCI.currentwg\_key   & T\_WK.theater\_key     & 0.545        \\
				T\_INCI.currentwg\_key   & T\_WK.subregion\_key    & 0.545        \\
				T\_INCI.currentwg\_key   & T\_WK.wkgrp\_key           & 0.545        \\
				T\_INCI.hwversion\_id     & T\_PROD.bl\_prod\_key    & 0.507         \\
				T\_INCI.createwkgrp\_key   & T\_WK.theater\_key     & 0.507         \\
				T\_INCI.createwkgrp\_key   & T\_WK.subregion\_key   & 0.507        \\
				T\_INCI.createwkgrp\_key   & T\_WK.wkgrp\_key       & 0.507         \\
				T\_PROD.item\_id        & X\_PRO.item\_id           & 0.468          \\
				T\_INCI.prod\_hw\_key    & T\_HW\_PROD.bl\_prod\_key & 0.463         \\
				T\_SUR.evalwkgrp\_key    & T\_WK.wkgrp\_key     & 0.433     \\ \bottomrule
			\end{tabular}
			
		\end{table}
		
	   \vspace{-0.4cm}
		\subsection{Evaluation based on Non-numerical Data}
		We evaluate our NEMA non-numerical algorithms based on TPM and Hashing on the non-numerical data in two parts as well. We use non-numerical ground truth data to evaluate the effectiveness of our algorithms. The matching results of all the other non-numerical field pairs are then described.
		
		\subsubsection{Evaluating of ground truth}
		The experiment for evaluating of non-numerical ground truths is shown here. There are 20 matched ground truth pairs which are annotated by humans. Additionally, 20 non-matched ground truth pairs are randomly selected and verified. Part of examples are shown in table \ref{tbl:NonNumGroundTruthExample}.
		
		% \usepackage{booktabs}
		\begin{table}[!th]
			\centering
			\caption{Examples of non-numerical ground truths}
			\label{tbl:NonNumGroundTruthExample}
			\scriptsize
			\begin{tabular}{p{0.3cm}p{2.4cm}p{2.4cm}p{1.6cm}}         %{@{}|lll|c|@{}}
				\toprule
				\textbf{No.} & \textbf{Table.field A}                                                   & \bf{Table.field B}                                                & \textbf{Matching class} \\ \midrule
				1                         & T\_PROD.item\_name                               & X\_INS.item\_name                 & 1              \\ 
				2                         & T\_COT.cpr\_country                        & T\_SITE.country                                  & 1              \\ 
				3                         & T\_CT.temp\_desc                & T\_PROD.item\_desc                     & 1              \\ 
				4                         & T\_CT.ctserv\_line                 & X\_SAH.servline\_name             & 1              \\ 
				5                         & T\_INCI.curr\_wg\_name    & T\_WK.wkgp\_name                     & 1              \\ 
				6                         & T\_CT.temp\_desc                & X\_SAH.temp\_name        & 1              \\ 
				7                         & T\_SITE.cust\_state                                 & X\_SAH.billto\_state           & 1              \\ 
				8                         & T\_CT.temp\_name                & X\_SAH.temp\_desc        & 1              \\ 
				9                         & T\_PROD.prod\_family                          & T\_HW\_PROD.family                       & 1              \\ 
				10                        & T\_PROD.prod\_family                          & T\_HW\_PROD.erp\_family                  & 1              \\ 
				1                         & T\_INCI.init\_gp\_name & T\_SITE.address                                  & 0              \\ 
				2                         & T\_DEFT.deft\_submitter                        & T\_SITE.email\_addr                           & 0              \\ 
				3                         & T\_COT.cpr\_country                        & T\_INCI.summary                  & 0              \\ 
				4                         & T\_SITE.address1                                & X\_PRO.prod\_family          & 0              \\ 
				5                         & T\_PROD.prod\_family                       & X\_SAH.hdrcust\_name             & 0              \\ 
				6                         & T\_INCI.tacpica\_ct      & T\_HW\_PROD.family                       & 0              \\ 
				7                          & T\_WK.wkgp\_desc                       & X\_PRO.physisn\_loc & 0              \\ 
				8                        & T\_SITE.county                                   & T\_SUR.batchcot\_name             & 0              \\ 
				9                        & T\_INCI.customersw\_ver    & T\_SITE.state                                    & 0              \\ 
				10                        & T\_OR\_LN.partsloc\_code  & X\_INS.item\_name                 & 0              \\ \bottomrule
			\end{tabular}
			
		\end{table}
		
		We first analyze the ground truth record-pairs and show the viability for the record pair similarity threshold $T_{rn}$.
		Table \ref{tbl:nonNumRecordSimExample} shows the record pair similarity scores of 9 different record pairs in a field pair (``T\_PROD.prod\_subgrp'', ``T\_HW\_PROD.platform''). The first 7 rows of pairs are matched record pairs that have higher similarity scores. The last 2 rows are not matched record pairs with lower score of 0.3333, which is lower than 0.4. It is conservative to capture the matching among different records with a similarity value 0.4 assigned to the threshold $T_{rn}$.

		\begin{table}[!th]
			\centering
			\caption{Sample of non-numerical record pairs }
			\label{tbl:nonNumRecordSimExample}
			\scriptsize
			\begin{tabular}{p{2.4cm}p{2.9cm}p{2cm}}    %{@{}|lll|@{}}
				\toprule
				{\textbf{T\_PROD.prod\_subgrp}} & \textbf{T\_HW\_PROD.platform} & \textbf{Record similarity} \\ \midrule
				c900 series                                              & c900 series                             & 1                 \\ 
				c2950 series                                             & c2916 series                            & 0.8      \\ 
				1601r series                                      & 1601 series                             & 0.775      \\ 
				css2950                                     & css2916                             & 0.667      \\ 
				C2960	 & 	C2960CX & 	0.577   \\ 
				C3560CX                  & C3560X  & 0.5      \\ 
				AIR35CE	& AIR35SE	& 0.4   \\ 
				
				\textbf{ts900}  & \textbf{cs900}    &\textbf{0.333}  \\ 
				\textbf{c800}                                              & \textbf{s800}                             & \textbf{0.333}               \\ \bottomrule
			\end{tabular}
			
		\end{table}
		
		%\begin{figure}[!th]
		%	\centering
		%	\includegraphics[width=3in,height=2.0in]{figures/nemath04_paperUse}
		%	\caption{Non-numerical matching ratios on ground truth}
		%	\label{fig:paperUseNonumericalMatchingRatio}
		%\end{figure}
		
\begin{figure}[!th]
	\centering
	\includegraphics[width=3.0in,height=1.7in]{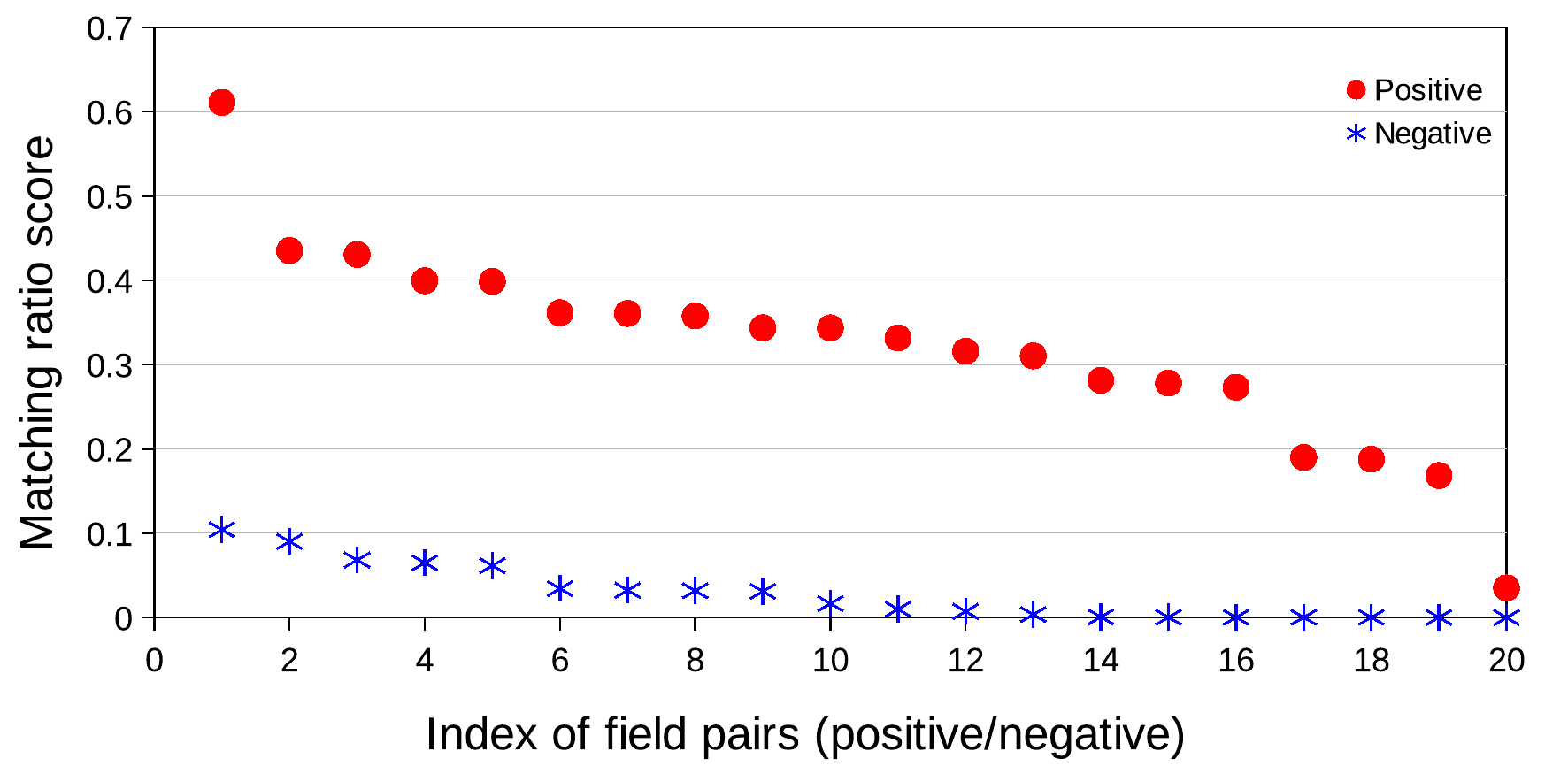}
	\caption{TPM-based matching ratio score}
	\label{fig:NEMA_NonNumericalGTFieldSimTPM}
\end{figure}

\begin{figure}[!th]
	\centering
	\includegraphics[width=3.15in,height=1.8in]{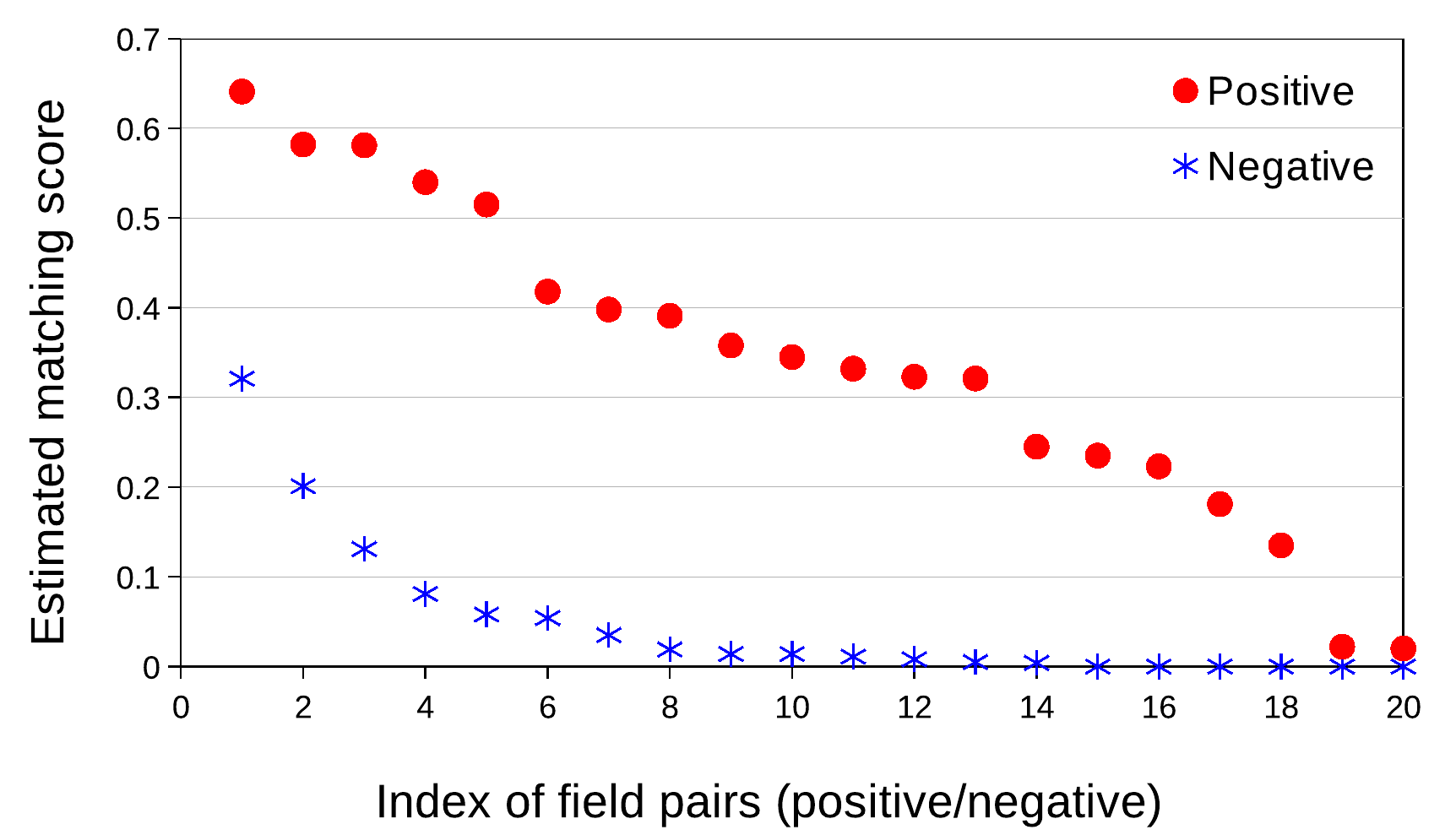}
	\caption{MH-LSH-based matching score}
	\label{fig:NEMA_NonNumericalGTFieldSimLSH}
\end{figure}

		We demonstrate the effectiveness of non-numerical algorithms based on TPM and MH-LSH by calculating matching ratio score based on TPM and estimated matching score based on MH-LSH on non-numerical ground truths. Figure \ref{fig:NEMA_NonNumericalGTFieldSimTPM} shows matching ratio scores of this ground truth data matching in a non-ascending order based on TPM. The matching ratio scores of almost all the matched pairs are above the non-matched pairs's. If we use the threshold 0.1 or select top-$20$ results from this, the accuracy can achieve about $95\%$, which shows the effectiveness of NEMA based on TPM. Figure \ref{fig:NEMA_NonNumericalGTFieldSimLSH} shows the matching scores of these ground truths in a non-ascending order based on MH-LSH. Although the decision boundary is not as good as the TPM-based result in Figure \ref{fig:NEMA_NonNumericalGTFieldSimTPM}, the accuracy can achieve 90\% when the threshold is 0.1 or top 19 results are selected.
		
%/home/fubao/workDir/ResearchProjects/CiscoWISH/CreateGraph/IdentifyExistingRelationship/Coma++Comparision/instance-level/COMACompare/nemaOut_GT/synthesisTH04
\vspace{-0.3cm}

%/ResearchProjects/CiscoWISH/CiscoWishMore/CreateGraph/GraphMatching/SpydeWks/minHash/output/nonnumericalOutput/
\vspace{-0.5cm}

\subsubsection{Top-$20$ similarity results}
\label{exper-nonnumericalTOpResult}

	There are 779 non-numerical fields in the large dataset. Considering that almost all the primary keys in a table are numerical fields, we do not consider primary key constraint matching method for non-numerical matching. The record similarity threshold  $T_{rn}$ is set to be 0.4 here based on the analysis of Table \ref{tbl:nonNumRecordSimExample}. The final top list of matching ratio scores are obtained based on TPM algorithm from all the non-numerical field pairs.

Table \ref{tbl:nonNumTop20MatchingRes} shows the top-$20$ results of field pair matching based on TPM. We can see that all the field pairs are matched pairs, and they are also confirmed by humans later,  which shows the effectiveness of NEMA based on TPM algorithm.

\begin{table}[!th]
	\centering
	\caption{Top-$20$ matching results of non-numerical field pairs}
			\label{tbl:nonNumTop20MatchingRes}
			\scriptsize
			\begin{tabular}{p{2.8cm}p{2.8cm}p{1.6cm}}     %{@{}lll@{}}
				\toprule
				\textbf{Table.field1}                                                 & \textbf{Table.field2}                                                 & \textbf{Matching ratio} \\ \midrule
				T\_INCI\_I2.currentwg\_key & T\_WK.wkgp\_name        & 0.637   \\ 
				T\_INCI\_I2.currentwg\_key & T\_WK.wkgp\_desc        & 0.632   \\ 
				T\_INCI.initwg\_name       & T\_WK.wkgp\_name        & 0.63    \\ 
				T\_INCI.initwg\_name      & T\_WK.wkgp\_desc         & 0.628   \\ 
				T\_WK.wkgmgr\_email                          & T\_SUR.eval\_email                         & 0.626   \\ 
				T\_INCI.creatorwg\_name              & T\_WK.wkgp\_name                           & 0.624   \\ 
				T\_INCI.creatorwg\_name              & T\_WK.wkgp\_desc                           & 0.622  \\ 
				T\_INCI.curr\_wg\_name       & T\_WK.wkgp\_name                           & 0.611   \\ 
				T\_INCI.curr\_wg\_name       & T\_WK.wkgp\_desc                          & 0.607   \\ 
				X\_SAH.billto\_state                 & T\_SITE.state                                          & 0.504 \\ 
				X\_SAH.billto\_state                 & T\_SITE.cust\_state                                    & 0.499  \\ 
				T\_INCI.initwg\_name    & T\_INCI\_I2.curr\_wg\_name      & 0.462    \\ 
				T\_INCI.initwg\_name       & T\_INCI\_I2.wkgrp\_name  & 0.457   \\ 
				T\_COT.cpr\_country                           & T\_SITE.country                                        & 0.453    \\ 
				T\_SITE.cust\_country                                  & T\_COT.cpr\_country                           & 0.453   \\ 
				T\_INCI\_I2.wkgrp\_name & T\_INCI.curr\_wg\_name       & 0.451 \\ 
				T\_COT.cpr\_country                           & T\_SITE.cust\_country                                  & 0.447   \\ 
				T\_INCI.curr\_wg\_name    & T\_INCI\_I2.curr\_wg\_name     & 0.442   \\ 
				T\_SITE.country                                        & T\_COT.cpr\_country                           & 0.44   \\ 
				T\_HW\_PROD.erpplatform                      & X\_SCDC.productsub\_grp            & 0.431   \\ \bottomrule
			\end{tabular}
		\end{table}
		
\vspace{-0.1cm}
\subsection{Evaluation of Classification-based Matching}		
As described in the section \ref{sec_classification}, we use SVM-based classifier to decide the matching of two field pairs. Here we show the numerical and non-numerical data testing result using cross-validation. We have generated 6000 synthetic numerical ground truth and non-numerical ground truth pairs based on the 60 numerical field pair and 40 non-numerical field pair, respectively. We run the whole experiment 20 times and obtain the average validation and testing accuracy which are shown in the table \ref{tab:svm_result}. It shows high accuracy of classification-based method to match field pairs with our proposed similarity metrics and also avoids the manual thresholds.

\begin{table}[!ht]
\caption{SVM validation and testing result on synthetic numerical and non-numerical ground truth}
\label{tab:svm_result}
\begin{tabular}{|l|l|l|}
\hline
Data/Average ACC   & Numerical & Non-numerical \\ \hline
Validation($10\%$)   & 0.995  & 0.953     \\ \hline
Testing($20\%$)      & 0.972  & 0.946      \\ \hline
\end{tabular}
\end{table}

\vspace{-0.2cm}
\subsection{Comparisons with baseline methods}
\vspace{-0.01cm}
  We compare our technique NEMA with baseline methods of COMA system \cite{engmann2007instance} and rule-based method Regex \cite{mehdi2012instance} here. COMA is a state of the art and popular hybrid matching tool and system supporting both schema-level and instance-level matching. Regex is an instance-level rule-based matching method based on regular expressions.  We test the numerical and non-numerical ground truth matching of NEMA and COMA in schema-level and instance-level and Regex in instance-level. Their accuracies and differences on matching results are compared and analyzed quantitatively.
		
\subsubsection{Comparison of Accuracy}
	We measure the accuracy and compare the COMA in the schema level and instance level and Regex in the instance level for numerical and non-numerical data matching. On the schema level matching, COMA uses the best field matching similarity "0" (which has no corresponding line in the COMA system) as a threshold in the schema-level matching. On the instance level matching, COMA has one similar instance-level matching that uses aggregated maximum record-wise similarities to obtain the final field pair similarities. The record-wise similarity is based on common similarity metrics such as edit distance \cite{ristad1998learning} and trigram \cite{angell1983automatic}. Edit distance is a technique to measure how dissimilar two strings are to one another by counting the minimum number of operations required to transform one string into the other. Trigram is a technique of splitting a string into triples of characters and comparing those to the trigrams of another string. The field matching similarity between two fields $A$ and $B$ in COMA is defined as follows:
	\begin{multline}
   sim(A,B) = \frac{1}{m+n}\cdot (\sum_{i=1}^{m}max_{j=1...,n}(sim(a_i,b_j)) + \\ \sum_{j=1}^{n}max_{i=1...,m}(sim(b_j,a_i)))
	\end{multline}
    Regex is a matching method based on regular expression by creating patterns from sampling instances of one field and then match against instances of another field to decide matching.
		
\begin{figure}[!th]
	\centering
\includegraphics[width=3.4in,height=2.2in]{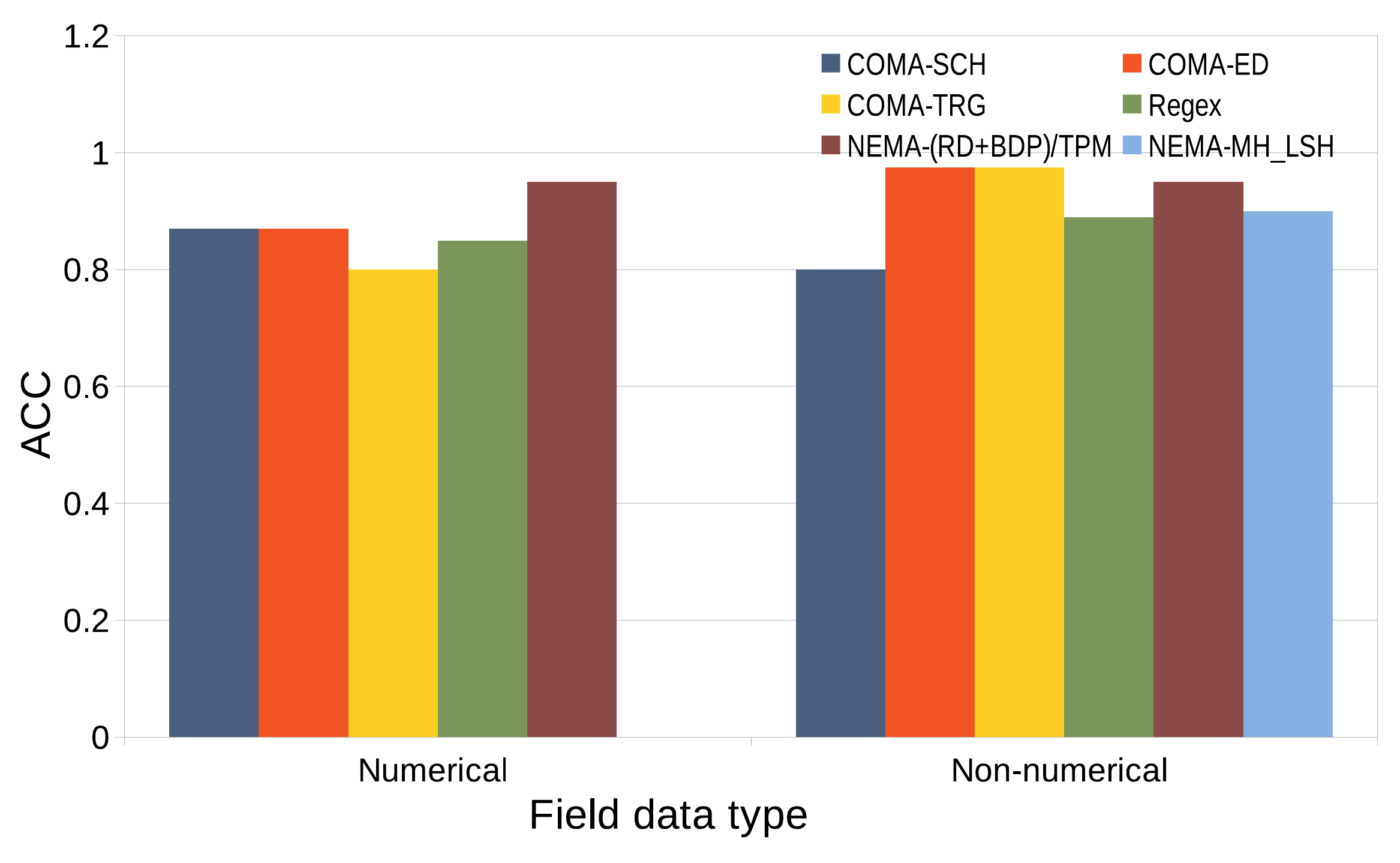}
\caption{Accuracy comparison}
 \label{fig:AccuracyComparison}
\end{figure}

	Figure \ref{fig:AccuracyComparison} shows the accuracy comparisons among COMA, NEMA and Regex. COMA-SCH is COMA with schema-level matching algorithm. COMA-ED and COMA-TRG means that COMA uses edit distance and trigram to measure the record similarity in instance-level matching, respectively. Regex is the matching based on regular expression in \cite{mehdi2012instance}. NEMA-(RD+BDP)/TPM is NEMA using RD and BDP in numerical data matching, and TPM in non-numerical data matching. NEMA-MH\_LSH indicates that NEMA uses minHash-locality sensitive hashing in non-numerical data matching, without MH-LSH bar shown for numerical matching.
		
	The accuracies of NEMA-(RD+BDP)/TPM in numerical and non-numerical data matching can be up to $95\%$, as high as COMA's non-numerical data matching, but having higher accuracy than COMA-SCH and Regex matching. For numerical matching, the accuracies of COMA-ED, COMA-TRG and Regex are $10\%$-$15\%$ lower than NEMA-(RD+BDP)/TPM because of the ineffectiveness to identify non-matched pairs of numerical ground truths. For non-numerical matching, the accuracies of COMA-ED and COMA-TRG are $2\%$ higher than NEMA-(RD+BDP)/TPM which is about $6\%$ higher than Regex. However, the field matching score of COMA is measured based on its general string similarity matching, which is not well applied to the network management database matching for record pair similarity requirements. A large number of pairs with high record similarities in COMA are not thought of as matches in the network management databases, which shows the usefulness of the NEMA non-numerical algorithm.
%ResearchProjects/CiscoWISH/CreateGraph/IdentifyExistingRelationship/Coma++Comparision
    \vspace{-0.4cm}
		
\subsubsection{Comparisons of Mismatched Examples}
		
  We further analyze the differences of COMA and NEMA in matching the ground truth field pairs. Table \ref{tbl:ExampleComparepair1} shows the field pairs in every row and its similarity scores by COMA and NEMA. These field pairs are found to be matched pairs by NEMA with relatively high similarity scores, but COMA shows no similarities with score 0. For COMA, the field names have very few common characters in spelling, even though the semantic commonality exists. NEMA does not rely on the inaccurate schema-level properties, but it uses the record instance for the decisions of field matching, which indirectly considers the semantic correspondences. If the record instances for some matched pairs are incomplete or missing, however, the similarity scores for these field pairs are also low. Table \ref{tbl:ExampleComparepair2} shows the two field pairs that have low similarities in NEMA. Although the field names in each pair express the same thing semantically, the record instances in the fields are actually incomplete and have very few in common between each other. However, in our databases, the missing or incomplete field pairs are very few compared to the large number of field pairs, which does not affect the overall performance for the numerical matching with NEMA.
		
		% \usepackage{booktabs}
		\begin{table}[!th]
			\centering
			\caption{Example of field pairs matched by NEMA, but not by COMA }
			\label{tbl:ExampleComparepair1}
			\scriptsize
			\begin{tabular}{p{1.9cm}p{1.9cm}p{1.76cm}p{1.76cm}}    %{@{}llll@{}}
				\toprule
				\textbf{Table.fieldA}                         & \textbf{Table.fieldB}                   & \textbf{COMA-ED} & \textbf{NEMA-TPM} \\ \midrule
				T\_INCI.ins\_site\_key  & T\_SITE.partysite\_id        & 0                     & 0.208              \\
				T\_OR\_HD.creator\_id         & T\_INCI.lastup\_by & 0                     & 0.643              \\
				T\_CT.temp\_desc & T\_PROD.item\_desc      & 0                     & 0.05                  \\ \bottomrule
			\end{tabular}
			
		\end{table}

		% \usepackage{booktabs}
		\begin{table}[!th]
			\centering
			\caption{Example of field pairs matched by COMA, but not by NEMA }
			\label{tbl:ExampleComparepair2}
			\scriptsize
			\begin{tabular}{p{1.8cm}p{2.4cm}p{1.3cm}p{1.4cm}}   %{@{}llll@{}}
				\toprule
				\textbf{Table.fieldA}                  & \textbf{Table.fieldB}                  & \textbf{COMA-ED} & \textbf{NEMA-TPM} \\ \midrule
				T\_INCI.bl\_cot\_key & T\_CT.bl\_cot\_key   & 0.750                  & 0.091              \\
				T\_SUR.bl\_surv\_key        & T\_SUR\_ANS.bl\_surv\_key & 0.76                  & 0.009              \\ \bottomrule
			\end{tabular}
			
		\end{table}
		
	Here we analyze the specific record pair examples of non-numerical instance-level matching. COMA uses standard edit distance and trigram to calculate the similarities of records, which is not quite suitable for the matching requirement of network management databases. Table \ref{ExampleCompare3NonnumericalInstance} below shows 9 examples of records pairs and three different kinds of similarities (NEMA-TPM, COMA-ED, COMA-TRG). The first 7 rows as one group are thought of as matched record pairs, the last two rows in the other group are non-matched record pairs. We can see from that the similarity of matched pairs based on COMA are quite similar around 0.7 for these two groups, from which is not easy to differentiate. While the similarities by NEMA have good differences (0.333 for non-matched pairs, 0.4 above for matched pairs). This further demonstrates that NEMA is more suitable for the network database matching.

		% \usepackage{booktabs}
		\begin{table}[!th]
			\centering
			\caption{Examples of record similarity comparisons}
			\label{ExampleCompare3NonnumericalInstance}
			\scriptsize
			\begin{tabular}{p{1.2cm}p{1.2cm}p{1.4cm}p{1.3cm}p{1.4cm}}  %{@{}lllll@{}}
				\toprule
				\textbf{Record 1}        & \textbf{Record 2}       & \textbf{NEMA-TPM} & \textbf{COMA-ED} & \textbf{COMA-TRG}  \\ \midrule
				c900 series     & c900 series    & 1                      & 1                    & 1              \\ 
				c2950 series    & c2916 series   & 0.8                    & 0.833             & 0.6            \\ 
				1601r series    & 1601 series    & 0.775                 & 0.909               & 0.738         \\ 
				css2950         & css2916        & 0.667                 & 0.714              & 0.6            \\ 
				C2960           & C2960CX        & 0.577                 & 0.6                  & 0.775        \\ 
				C3560CX         & C3560X         & 0.5                    & 0.833               & 0.671         \\ 
				AIR35CE         & AIR35SE        & 0.4                    & 0.857               & 0.6            \\ 
				\textbf{c800}   & \textbf{s800}  & \textbf{0.333}         & \textbf{0.75}        & \textbf{0.5}   \\ 
				\textbf{ts 900} & \textbf{cs900} & \textbf{0.333}         & \textbf{0.8}         & \textbf{0.667} \\ \bottomrule
			\end{tabular}
			
		\end{table}

\subsubsection{Comparison of Efficiency}
\begin{figure}[!th]
	\centering
\includegraphics[width=3.5in,height=2.3in]{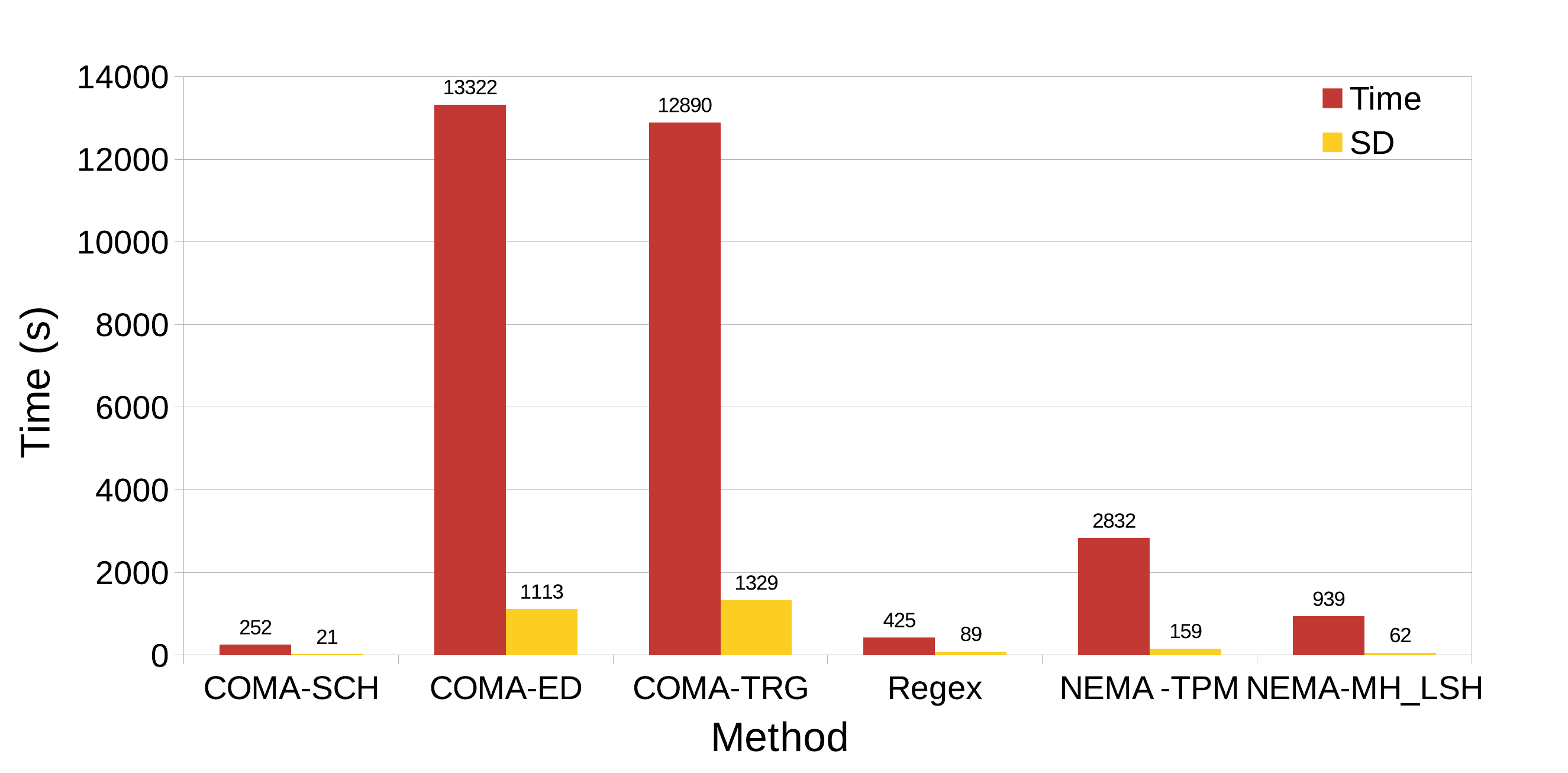}
	\caption{Efficiency comparison}
	\label{fig:SpeedupComparisonnNonnumerical}
\end{figure}

Considering the expensive time consumption for non-numerical field pair matching, we test the efficiency based on the whole non-numerical ground truth. We run the experiment 20 times on the same machine with the same data to calculate the average computation time and standard deviation (SD) without data loading time. Figure \ref{fig:SpeedupComparisonnNonnumerical} shows the total computation time spent for COMA-SCH, COMA-ED, COMA-TRG, Regex, NEMA-TPM and NEMA-MH\_LSH. COMA-SCH and Regex using schema information only are fastest among all methods, but the accuracies are the lowest. Except from from COMA-SCH and Regex, COMA-ED is the slowest, taking about 12,000 seconds. NEMA-TPM takes about 2500 seconds, about 5 times speedup than COMA-ED. NEMA-MH\_LSH takes 860 seconds, which is about 14 times faster than COMA-ED. NEMA-TPM reaches the best trade-off between accuracy and efficiency among these algorithms.
        
		% Note that the IEEE does not put floats in the very first column
		% - or typically anywhere on the first page for that matter. Also,
		% in-text middle ("here") positioning is typically not used, but it
		% is allowed and encouraged for Computer Society conferences (but
		% not Computer Society journals). Most IEEE journals/conferences use
		% top floats exclusively. 
		% Note that, LaTeX2e, unlike IEEE journals/conferences, places
		% footnotes above bottom floats. This can be corrected via the
		% \fnbelowfloat command of the stfloats package.
		
\section{Related Work}
\label{RelatedWork}
	The structured data matching is an old and important research topic but unsolved and ever-growing problem, which has a wide range of applications in database integration, migration, semantic query, etc. \cite{castano2018matching}. In survey papers \cite{rahm2001survey, shvaiko2005survey}, the authors proposes a solution taxonomy differentiating between element and structure level, schema and instance level, language and constraint-based matching techniques. Furthermore, P. Shivaiko et al. \cite{shvaiko2013ontology} reviews the state-of-the-art matching systems which were based on strings, structure, data instance and semantics matching techniques using different schema formats such as database, XML, OWL, RDFS, etc. In database schema matching, previous common matching systems in schema-level are introduced in several prototypes such as  Similarity Flooding (SF) \cite{melnik2002similarity}, Coma \cite{aumueller2005schema}, etc.
		
SF \cite{melnik2002similarity} is a matching algorithm that models two structured columns to be compared as two directed labeled graphs. It makes use of field data, key properties and the string-based alignment (prefix and suffix test) to obtain the alignments between two nodes of the graph. The similarity is calculated from similar nodes to adjacent neighbors through propagation. Our NEMA only relies on the data instance values to infer the matching of fields, which does not utilize the structured properties and data types. However, SF uses a metric for matching quality based on the intended matching results, which is similar to our accuracy metric based on top-$k$ results.

Coma \cite{aumueller2005schema} is a composite matching system providing extensible library and framework for combining obtained results. It contains mainly 6 elementary matchers using string-based techniques, 5  hybrid matchers using names and structural paths, and one reuse-oriented matcher based on previous matching results . The composite matcher effectively improves the match quality over single matchers using the default combination strategy. Compared to SF, the overall average matching quality are the best among them \cite{aumueller2005schema}. The extended version Coma++ \cite{engmann2007instance} utilizes the shared taxonomy and pivot schema to further improve the overall matching quality.  In our evaluation, we compare with the Coma++ method using the default combination strategy and find our technique NEMA overall outperforms than COMA in schema-level matching.
		
Except from the previous matching approaches using field and structural information matching, data instanced-based approaches \cite{nottelmann2007information, madhavan2005corpus, chen2012mining, yang2008effective} use the similarity metric or machine learning or rule-based methods to determine the similarity of fields. In \cite{madhavan2005corpus}, the authors utilize a corpus that contains schema and mappings between some schema pairs, and learn the constraints from schema statistics to help generate more matching pairs. In \cite{chen2012mining}, the authors uses the mutual information of statistics to measures the similarity of schema instances between two columns to decide the matching, which shows an effective method based on instances. Also, the authors in \cite{qian2012sample} propose a new sample-driven approach which enables the end-users to easily construct their own data to match the source and target schema. \cite{mehdi2012instance} proposes a rule-based method by creating regular expression pattern to match against columns. \cite{yang2008effective} uses matching-learning technique of training neural networks for getting candidate pairs and then filters the pairs with a rule-based algorithm. Our NEMA uses proposed similarity metrics as features to train a SVM to classify for matching effectively and efficiently. COMA \cite{engmann2007instance} proposes two instance-level matching methods based on the constraint of instance data and the content-based matching to measure field matching. The constraint-based method relies on the general, numerical and pattern constraint which has specific limitation to the specific data which is not suitable for the network databases. The content-based matching depends on the aggregation of similarity scores of instance contents and it is kind of similar to our NEMA technique on content-based similarity measurement, which are compared with NEMA in the experimental evaluations.
		
To sum up, most of currently popular matching approaches and systems focus on schema-level information matching. The data instances level matching approaches using field record values are mostly based on some statistical models and machine learning from corpus. We further explore the database instance matching by comparing field records using different metrics and propose effective and overall matching algorithms considering the characteristic of network database matching for a graph database construction for efficient data query, analysis and management.

\section{Conclusion}
\label{Conclusion}
In this paper we propose a systematic technique NEMA to match databases for network management. Different from previous database matching approaches, we design a technique to match numerical and non-numerical fields in instance-level respectively, which can effectively be integrated into a graph database for network management and analysis. For numerical matching, we propose range difference similarity and bucket dot product similarity metrics. For non-numerical matching, we design top priority match metric and also propose applying minHash-locality sensitive hashing algorithm, which reduces the matching time for large databases. To address the drawback of manual thresholds, an effective classification-based method is also proposed based on the proposed similarity metrics. The results of NEMA are demonstrated to be promising with better efficiency and comparable accuracy than the baseline algorithms.
		
With the explosion of big data and popularity of distributed graph processing systems, this work has the potential to significantly reduce the human work involving identifying the matching fields for a large graph database construction and also be applied for large-scale data matching. A majority of partial matching pairs can be found by our matching algorithms which are not easily detected by humans.

%		\section*{Acknowledgment}
%		The authors would like to thank...
		
		\ifCLASSOPTIONcaptionsoff
		\newpage
		\fi
		
		
		
		%\bibliographystyle{IEEEtran}
		\bibliographystyle{abbrv}
		\bibliography{nema} 
		
	\vspace{-1.5cm} 
		\begin{IEEEbiography}[{\includegraphics[width=1in,height=1.25in,clip,keepaspectratio]{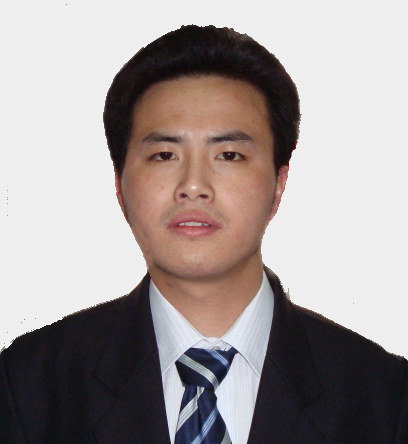}}]{Fubao Wu} received his Bachelor degree in Electronic Information Engineering from Northeastern 
University, China in 2008 and Master of Engineering in Electronic Science and Technology from University of Science and Technology of China in 2011. He is currently pursuing his Ph.D. degree
in Electrical and Computer Engineering at University of Massachusetts, Amherst. His research interests are data analytics, graph analytics and video analytics.
		\end{IEEEbiography}
		
	\vspace{-1.5cm} 
		\begin{IEEEbiography}[{\includegraphics[width=1in,height=1.25in,clip,keepaspectratio]{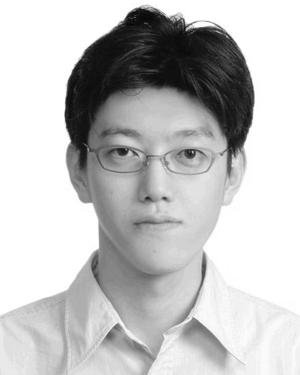}}]{Han Hee Song}
			was Lead Data Scientist at Cisco
Systems. He was Principal Data Scientist at Symantec Corp. and Senior Member of Technical Sta at CTO office of Narus, Inc. His research focuses on privacy analysis of mobile users and protective measures for cyber terrorism. Dr. Song received a Ph.D. and M.A. in Computer Science from the University of Texas at Austin in 2011 and 2006, respectively.
He received a B.S. from Yonsei University, Seoul, Korea in 2004.
\end{IEEEbiography}
		
		%\newpage
	\vspace{-1.5cm} 
		\begin{IEEEbiography}[{\includegraphics[width=1in,height=1.25in,clip,keepaspectratio]{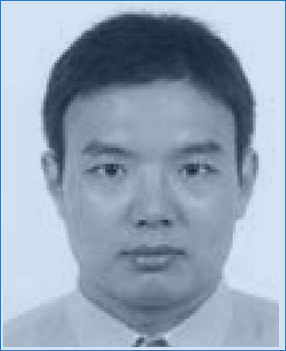}}]{Jiangtao Yin}
			received the Ph.D. degree in Electrical and Computer Engineering from the University of Massachusetts at Amherst, in 2016. He is currently working at Palo Alto Networks. His current research interests include cloud computing, large-scale data processing, stream processing, and distributed systems.
		\end{IEEEbiography}
	\vspace{-1.5cm} 
		\begin{IEEEbiography}[{\includegraphics[width=1in,height=1.25in,clip,keepaspectratio]{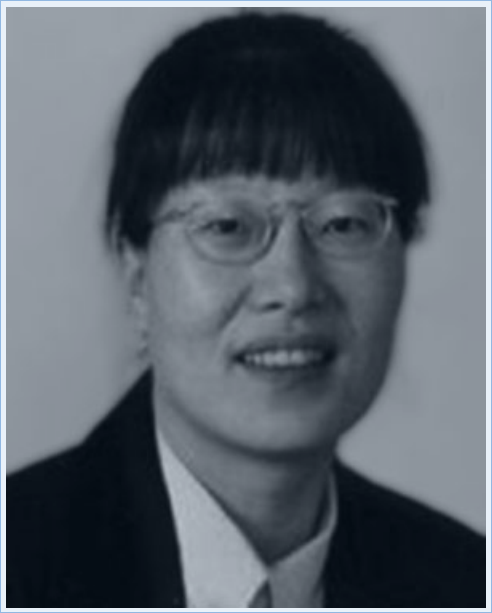}}]{Lixin Gao}
			received the Ph.D. degree in computer science from the University of Massachusetts at Amherst, in 1996. Now she is a professor of electrical and computer engineering with the University of Massachusetts at Amherst. Her research interests include social networks, Internet routing, network virtualization and cloud computing. Between May 1999 and January 2000, she was a visiting researcher in AT\&T Research Labs and DIMACS. She was an Alfred P. Sloan fellow between 2003-2005 and received an NSF CAREER Award in 1999. She won the best paper award from IEEE INFOCOM 2010 and ACM SoCC 2011, and the test-of-time award in ACM SIGMETRICS 2010. She received the Chancellors Award for Outstanding Accomplishment in Research and Creative Activity in 2010. She is a fellow of the IEEE and ACM.
		\end{IEEEbiography}
	\vspace{-1.1cm} 
		\begin{IEEEbiography}[{\includegraphics[width=1in,height=1.25in,clip,keepaspectratio]{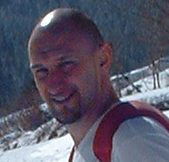}}]{Mario Baldi}
			 is Director of Technology at Cisco Systems and Associate Professor at Politecnico di Torino. He was Data Scientist Director at Symantec Corp., Inc., Principal Member of Technical Sta at Narus, Inc., Principal Architect at Embrane, Inc.; Vice Dean of the PoliTong Sino-Italian Campus at Tongji University, Shanghai; Vice President for Protocol Architecture at Synchrodyne Networks, Inc., New York. Through his research, teaching and professional activities, Mario Baldi has built considerable knowledge and expertise in big data analytics, next generation network data analysis, internetworking, high performance switching, optical networking, quality of service, multimedia networking, trust in distributed software execution, and computer networks in general.
		\end{IEEEbiography}
	\vspace{-1.1cm} 
		\begin{IEEEbiography}[{\includegraphics[width=1in,height=1.25in,clip,keepaspectratio]{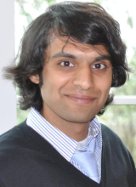}}]{Narendra Anand} worked as a Research Scientist at Cisco Systems as a part of the Forward-looking Analytics, Software, and Technologies (FAST) program. He received his Ph.D. degree in May of 2015 at Rice University in the Department of Electrical and Computer Engineering. His research focuses on the design and implementation of novel, cross-layer, Multi-user Multiple-Input-Multiple-Output (MU-MIMO) communication protocols.
		\end{IEEEbiography}
		\vfill
		
		
		%\vfill
		

	\end{document}